\documentclass{nature}

\usepackage[top=0.8in, bottom=1in, left=0.8in, right=0.8in]{geometry}

\usepackage{graphicx}
\usepackage{amsmath}
\usepackage{amssymb}
\usepackage{threeparttable}
\usepackage{caption}
\usepackage{bibunits}
\usepackage{color}
\usepackage{adjustbox}
\usepackage{url}
\usepackage{soul}

\defaultbibliography{MasterBiblio.bib}
\defaultbibliographystyle{naturemag}

\title{Normal, Dust-Obscured Galaxies in the Epoch of Reionization}

\author{
Y. Fudamoto$^{1,2,3}$, 
P. A. Oesch$^{1,4}$,
S. Schouws$^{5}$,
M. Stefanon$^{5}$,
R. Smit$^{6}$,
R. J. Bouwens$^{5}$,
R. A. A. Bowler$^{7}$,
R. Endsley$^{8}$,
V. Gonzalez$^{9,10}$,
H. Inami$^{11}$,
I. Labbe$^{12}$,
D. Stark$^{8}$,
M. Aravena$^{13}$,
L. Barrufet$^{1}$,
E. da Cunha$^{14,15}$,
P. Dayal$^{16}$,
A. Ferrara$^{17}$,
L. Graziani$^{18,20,27}$,
J. Hodge$^{5}$,
A. Hutter$^{16}$,
Y. Li$^{21,22}$,
I. De Looze$^{23,24}$,
T. Nanayakkara$^{12}$,
A. Pallottini$^{17}$,
D. Riechers$^{25}$,
R. Schneider$^{18,19,26,27}$,
G. Ucci$^{16}$,
P. van der Werf$^{5}$,
C. White$^{8}$
\vspace{8pt}}

\begin{document}
\maketitle

\begin{affiliations}
\small
 \item Department of Astronomy, University of Geneva, 51 Ch. Pegasi, 1290 Versoix, Switzerland
 \item  Research Institute for Science and Engineering, Waseda University, 3-4-1 Okubo, Shinjuku, Tokyo 169-8555, Japan
 \item National Astronomical Observatory of Japan, 2-21-1, Osawa, Mitaka, Tokyo, Japan
 \item Cosmic Dawn Center (DAWN), Niels Bohr Institute, University of Copenhagen, Jagtvej 128, K\o benhavn N, DK-2200, Denmark
 \item Leiden Observatory, Leiden University, PO Box 9500, 2300 RA Leiden, The Netherlands
 \item Astrophysics Research Institute, Liverpool John Moores University, 146 Brownlow Hill, Liverpool L3 5RF, UK
 \item Sub-department of Astrophysics, The Denys Wilkinson Building, University of Oxford, Keble Road, Oxford, OX1 3RH, UK
 \item Steward Observatory, University of Arizona, 933 N Cherry Ave, Tucson, AZ 85721, USA
 \item Departmento de Astronomia, Universidad de Chile, Casilla 36-D, Santiago 7591245, Chile
 \item Centro de Astrofisica y Tecnologias Afines (CATA), Camino del Observatorio 1515, Las Condes, Santiago 7591245, Chile
 \item Hiroshima Astrophysical Science Center, Hiroshima University, 1-3-1 Kagamiyama, Higashi-Hiroshima, Hiroshima 739-8526, Japan
 \item Centre for Astrophysics \& Supercomputing, Swinburne University of Technology, PO Box 218, Hawthorn, VIC 3112, Australia
 \item Nucleo de Astronomia, Facultad de Ingenieria y Ciencias, Universidad Diego Portales, Av. Ejercito 441, Santiago, Chile
 \item International Centre for Radio Astronomy Research, University of Western Australia, 35 Stirling Hwy, Crawley, WA 6009, Australia
 \item ARC Centre of Excellence for All Sky Astrophysics in 3 Dimensions (ASTRO 3D)
 \item Kapteyn Astronomical Institute, University of Groningen, PO Box 800, NL-9700 AV Groningen, the Netherlands
 \item Scuola Normale Superiore, Piazza dei Cavalieri 7, 56126 Pisa, Italy
 \item Dipartimento di Fisica, Sapienza, Universita di Roma, Piazzale Aldo Moro 5, I-00185 Roma, Italy
 \item INAF/Osservatorio Astronomico di Roma, via Frascati 33, 00078 Monte Porzio Catone, Roma, Italy
 \item INAF/Osservatorio Astrofisico di Arcetri, Largo E. Femi 5, I-50125 Firenze, Italy
 \item Department of Astronomy \& Astrophysics, The Pennsylvania State University, 525 Davey Lab, University Park, PA 16802, USA
 \item Institute for Gravitation and the Cosmos, The Pennsylvania State University, University Park, PA 16802, USA
 \item Sterrenkundig   Observatorium,   Ghent   University,   Krijgslaan 281 - S9, 9000 Gent, Belgium
 \item Dept. of Physics \& Astronomy, University College London,Gower Street, London WC1E 6BT, UK
 \item Cornell University, 220 Space Sciences Building, Ithaca, NY 14853, USA
 \item Sapienza School for Advanced Studies, Viale Regina Elena 291, 00161 Roma Italy
 \item INFN, Sezione di Roma 1, P.le Aldo Moro 2, 00185 Roma, 
\end{affiliations}

\begin{bibunit}

\begin{abstract}
Over the past decades, rest-frame ultraviolet (UV) observations have provided large samples of UV luminous galaxies at redshift (z) greater than 6\cite{Madau2014,Bouwens15aLF,Ono18}, during the so-called epoch of reionization. 
While a few of these UV identified galaxies revealed significant dust reservoirs\cite{Watson15,Hashimoto19,Tamura19,Bakx20}, very heavily dust-obscured sources at these early times have remained elusive. They are limited to a rare population of extreme starburst galaxies\cite{Riechers2013,Strandet17,Marrone2018,Dudzeviciute20,Riechers2020}, and companions of rare quasars\cite{Decarli2017,Mazzucchelli2019}. These studies conclude that the contribution of dust-obscured galaxies to the cosmic star formation rate density at $\mathbf{z>6}$ is sub-dominant. Recent ALMA and Spitzer observations have identified a more abundant, less extreme population of obscured galaxies at $\mathbf{z=3-6}$\cite{Wang2019,Williams2019}. However, this population has not been confirmed in the reionization epoch so far.
Here, we report the discovery of two dust-obscured star forming galaxies at $\mathbf{z=6.6813\pm0.0005}$ and $\mathbf{z=7.3521\pm0.0005}$. These objects are not detected in existing rest-frame UV data, and were only discovered through their far-infrared [CII] lines and dust continuum emission as companions to typical UV-luminous galaxies at the same redshift. The two galaxies exhibit lower infrared luminosities and star-formation rates than extreme starbursts, in line with typical star-forming galaxies at $\mathbf{z\sim7}$. 
This population of heavily dust-obscured galaxies appears to contribute 10-25 per cent to the $\mathbf{z>6}$ cosmic star formation rate density.
\end{abstract}

As part of the ongoing ALMA large program REBELS (Reionization-Era Bright Emission Line Survey), we are observing 40 UV-luminous primary targets at $z>6.5$\cite{Bouwens21}.
Among these targets are REBELS-12 and REBELS-29. When inspecting the ALMA data cube of these two sources, we identified two additional emission line neighbors.

\begin{figure*}[h!]
  \begin{center}
  \includegraphics[width=.85\linewidth]{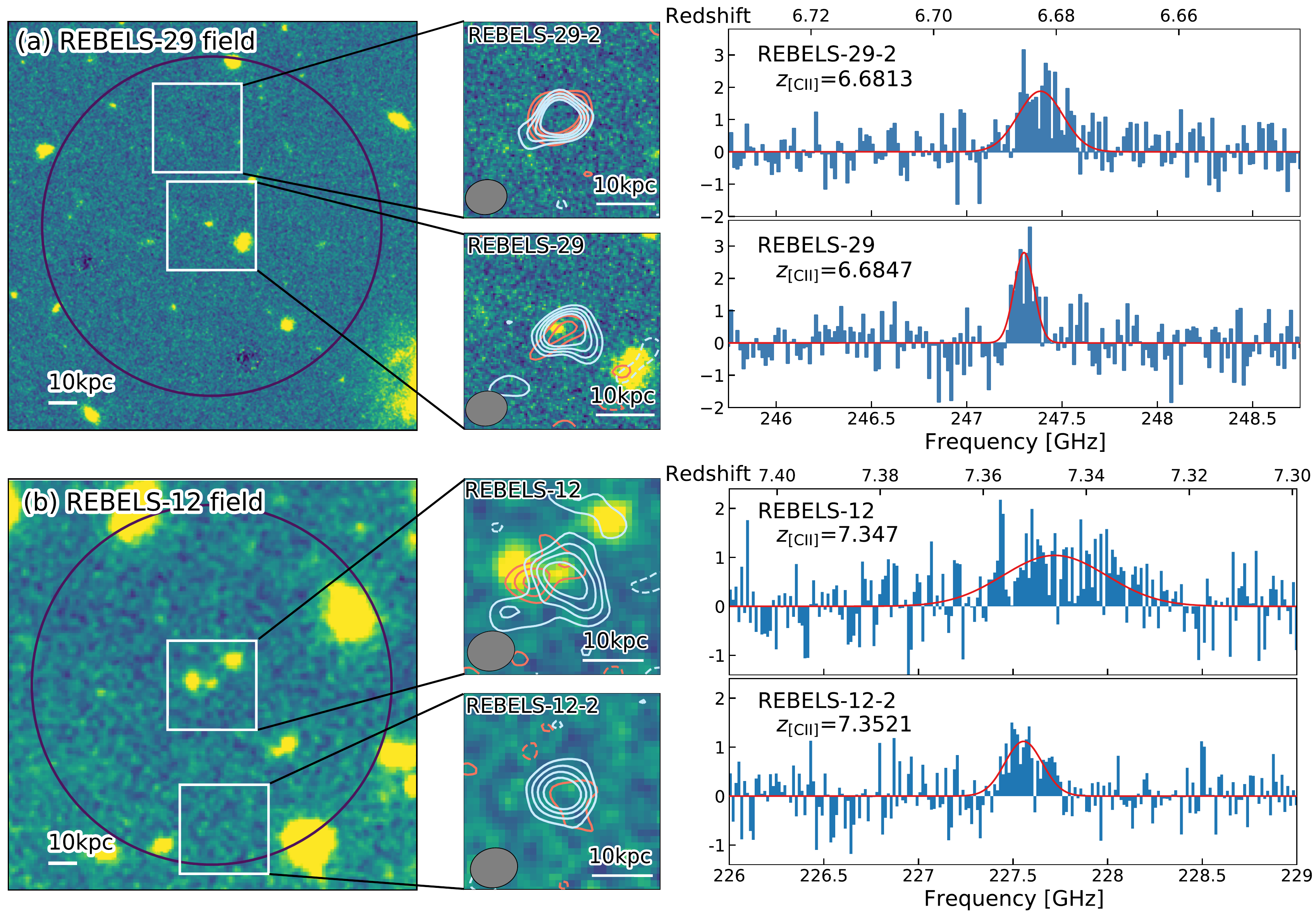}
  \end{center}
    \vspace{-0.6truecm}  
    \caption{\small\textbf{[CII] $\mathrm{\mathbf{158\,\mu m}}$ line and dust emission detections.} {\bf a:} REBELS-29 field at $z\sim6.68$. {\bf b:} REBELS-12 field at $z\sim7.35$. Background images are HST F140W and VIDEO J-band, respectively. Black circles show the half power beam widths of each ALMA pointing ($\sim13^{\prime\prime}$ radii), and white squares indicate $6^{\prime\prime}.5$ wide boxes that are shown in the middle panels. White horizontal bars correspond to 10 proper kpc. Solid red and light blue contours show $2\sigma$ to $5\sigma$ levels (and $-5\sigma$ to $-2\sigma$ for dashed contours) for the continuum and [CII] moment-0 maps, respectively. The continuum subtracted [CII] spectra are shown at the native velocity resolution of $\mathrm{20\,km/s}$. The two sources REBELS-29-2 and REBELS-12-2 were found serendipitously as companions to the central, UV-luminous targets, with emission lines at almost exactly the same frequencies as the central targets, accompanied with dust continuum emission at the same location. Their spatial and spectral proximity, and absence in optical/NIR images confirms these companions as unexpected, dusty star forming sources in the epoch of reionization.}
   \label{fig:stamps}
    \vspace{-12pt}
\end{figure*}

The primary targets of our ALMA observing program, REBELS-12 and REBELS-29, represent some of the most UV-luminous galaxies in this redshift range, and have $M_\mathrm{UV}=-22.5\pm0.3$ and $-22.2\pm0.1.$ They were originally identified with a photometric redshift of $z_\mathrm{phot}=6.82_{-0.11}^{+0.13}$ and $z_\mathrm{phot}=7.40_{-0.21}^{+0.15}$, respectively, based on deep ground based optical and near-infrared (NIR) data.
The ALMA observations were carried out on 24 and 29 November 2019, targeting the singly ionized carbon emission line, [CII] $158\,\mathrm{\mu m}$ and dust continuum emission with a frequency coverage of the vast majority of the photometric redshift probability distribution.
The ALMA observations reached  emission line sensitivities of $0.19\,\mathrm{mJy\,beam^{-1}}$ and $0.16\,\mathrm{mJy\,beam^{-1}}$ per $100\,\mathrm{km/s}$ spectral element for REBELS-29 and REBELS-12, respectively. This resulted in clear [CII] emission line detections of both sources, at frequencies perfectly consistent with the photometric redshift estimations (see Figure~1 and Extended Data Figure~1).
The integrated flux densities of these lines are $0.44\,\mathrm{Jy\,km\,s^{-1}}$ and $1.20\,\mathrm{Jy\,km\,s^{-1}}$ corresponding to point source detection significances of $9.2\,\sigma$ and $6.3\,\sigma$, respectively. These lines yield a spectroscopic redshift of $z=6.6847\pm0.0002$ and $z=7.347\pm0.001$ for REBELS-29 and REBELS-12, respectively.

\begin{figure*}[th!]
    \centering
    \includegraphics[width=0.88\textwidth]{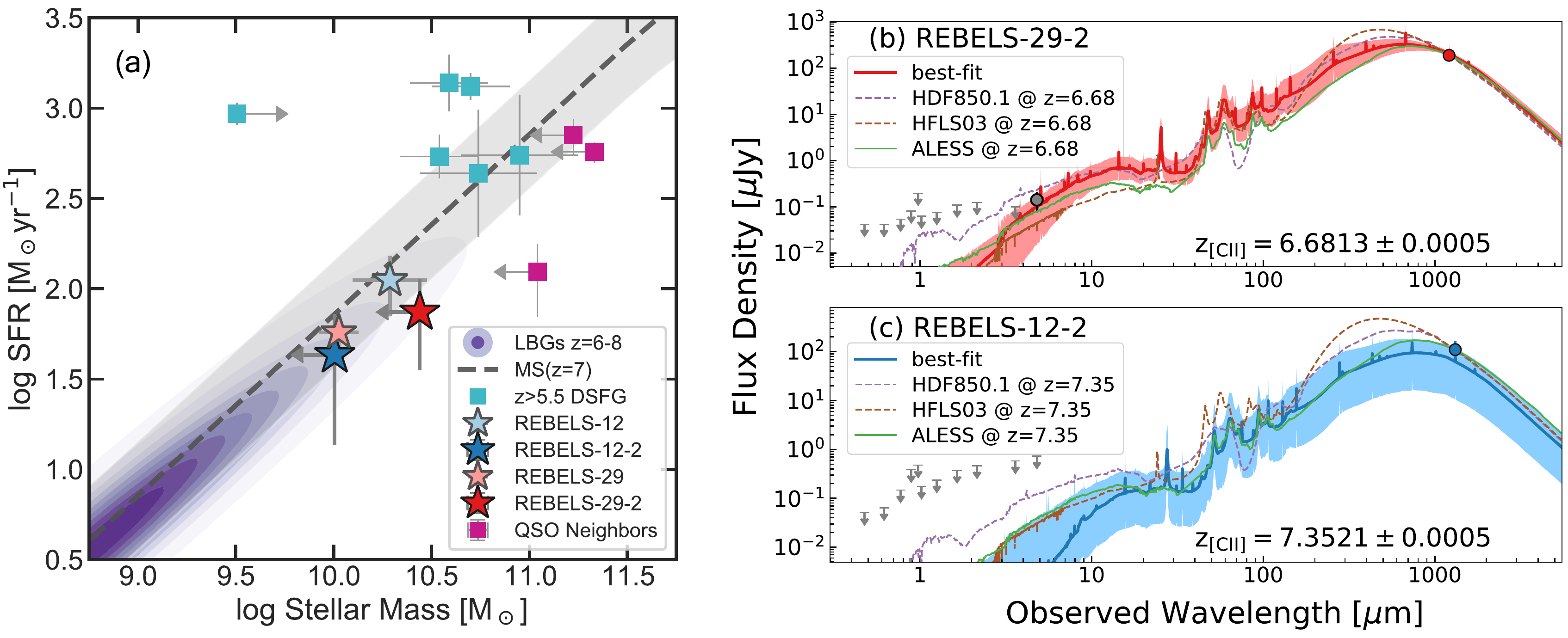}
    \caption{\small \textbf{Estimated properties of REBELS-29-2 and REBELS-12-2.} {\bf a:} Star-formation rate as a function of stellar mass for $z>5.5$ galaxies. Background contours show the distribution of faint LBGs at $z=6-8$. The dashed line and shaded region indicates the star forming main-sequence as measured up to $z\sim4$ and extrapolated to $z=7$\cite{Schreiber2015}. Cyan squares represent previously identified dusty star-forming galaxies (DSFGs) at $z\sim5.7-6.9$. Purple squares show dusty galaxies found as companions of $z\sim6.1-6.5$ quasars, which also remained undetected in rest-UV observations. Their mass limits are estimated from dynamical masses\cite{Decarli2017,Mazzucchelli2019}. Error bars correspond to $1\sigma$ uncertainties. The UV-bright galaxies (REBELS-12 and REBELS-29), and the serendipitous, dusty galaxies (REBELS-12-2 and REBELS-29-2) are shown as blue and red stars (using SFR$_\mathrm{IR}$). The SFRs and stellar mass limits of the newly identified galaxies are lower than the majority of $z>5.5$ DSFGs and quasar companions at these redshifts. {\bf b} and {\bf c}: Photometric constraints and SEDs of REBELS-29-2 and REBELS-12-2, respectively. 2$\sigma$ upper limits are shown for non-detections (gray arrows). Red and blue solid lines show the median posterior SEDs together with their $68\,\%$ confidence contours. For comparison, SEDs of dusty star-bursting galaxies normalized at the ALMA continuum fluxes are indicated in solid green (the average SED of ALESS galaxies\cite{Swinbank2014}), dashed brown (HFLS03\cite{Riechers2013}), and dashed purple lines (HDF850.1\cite{Walter2012}). The newly found dusty sources as companions of normal star-forming galaxies suggest that dusty, lower-luminosity versions of DSFGs exist at $z=6-8$ in larger number than previously assumed
    }
    \label{fig:MS_SED}
\end{figure*}

The [CII] line integrated maps of these galaxies revealed two strong, serendipitous emission lines at $\sim7.4$ and $\sim11.5$ arcsec offset from the primary targets, corresponding to $40\,\mathrm{pkpc}$  and $58\,\mathrm{pkpc}$, respectively.
The continuum subtracted spectra then confirmed that the additional emission lines emerge at almost exactly the same frequency as the central targets (velocity offsets of $110\,\mathrm{km/s}$ and $210\,\mathrm{km/s}$ for REBELS-29 and REBELS-12, respectively; see Right Panels of Figure~1).
Hereafter, we call these additional line sources REBELS-12-2 and REBELS-29-2.

After correcting for the primary beam attenuation, these emission lines have integrated flux densities of $0.781\,\mathrm{Jy\,km\,s^{-1}}$ and $0.581\,\mathrm{Jy\,km\,s^{-1}}$ corresponding to a detection significance of $9.7\,\sigma$ and $6.8\,\sigma$ in the moment-0 maps for REBELS-29-2 and REBELS-12-2, respectively. Additionally, REBELS-29-2 has a clear ($9.2\,\mathrm{\sigma}$) dust continuum detection, while only a tentative ($2.8\,\mathrm{\sigma}$) continuum signal is found for REBELS-12-2.
These measurements correspond to flux densities of $192\pm25\,\mathrm{\mu Jy}$ and $110\pm52\,\mathrm{\mu Jy}$ after applying primary beam and cosmic microwave background (CMB) corrections, respectively (see Extended Data Table 1).

Given the significance of the emission line detections and the number of independent beams in the moment-0 maps, the probability of a random Gaussian noise fluctuation is negligible. 
Furthermore, the co-spatial continuum signals confirm that these blind line detections are real.
In the Method section, we estimate the probability of finding an unassociated, random emission line almost exactly at the same frequency as a primary one in an ALMA data cube to be extremely small ($p<6\times10^{-4}$). Therefore, this strongly suggests that these serendipitous detections stem from neighboring sources sitting in the same environment as the primary targets. The [CII] $158\,\mathrm{\mu m}$ redshifts of REBELS-29-2 and REBELS-12-2 are thus $z=6.6813\pm0.0002$ and $z=7.3521\pm0.0005$, respectively.

REBELS-29-2 and REBELS-12-2 are covered by deep optical/NIR observations, corresponding to rest-frame UV/optical wavelengths.
However, neither source shows any optical counterpart (Figure~1 and Extended Data Figure~2) except for a tentative ($\sim2\,\sigma$) Spitzer IRAC $4.5\,\mathrm{\mu m}$ detection of REBELS-29-2.
These non-detections in rest-frame UV/optical bands indicate that these serendipitously found [CII] line emitters are heavily dust-obscured galaxies.

SFRs are estimated based on the UV and FIR luminosities. However, due to the non-detections in rest-UV data we can only provide limits on the UV-based SFRs, for which we find $\mathrm{SFR_{UV}}<2$ and $<14.7\,\mathrm{M_{\odot}\,yr^{-1}}$ (at 2$\sigma$); the IR-based SFRs from the ALMA dust continuum amount to $\mathrm{SFR_{IR}}=74.3^{+47.0}_{-31.2}$ and $43.1^{+37.0}_{-22.7}\,\mathrm{M_{\odot}\,yr^{-1}}$ for REBELS-29-2 and REBELS-12-2, respectively.
We hereafter use the SFRs directly derived from FIR luminosities as our fiducial SFRs (for more information, see Methods).

Given the optical non-detections of REBELS-29-2 and REBELS-12-2, also their stellar masses cannot be accurately constrained. We only derive upper limits from spectral energy distribution (SED) fitting (see Methods). The estimated SFRs and stellar mass limits are consistent with normal $\sim$L$_*$ Lyman Break galaxies (LBGs) at $z>6$ selected from rest-frame UV data such as the REBELS primary target sample, which define the so-called main-sequence of star-formation\cite{Schreiber2015} (Figure~2).

\begin{table*}[h]
    \centering
    \footnotesize
    \begin{tabular}{lccccccc}
    \multicolumn{8}{l}{\textbf{Table 1: Summary of Galaxy Properties}} \\[0.1cm]
    \hline
        Galaxy Name & Redshift$^{\dagger}$\tnote{dagger} & RA & Dec & Stellar Mass & $\mathrm{SFR_{UV}}$ & $\mathrm{SFR_{IR}}$$^{\dagger\dagger}$\tnote{ddagger} & $\mathrm{SFR_{[CII]}}$ \\
         & & (deg) & (deg) & ($10^{9}\,M_{\odot}$) &  ($M_{\odot}\,\mathrm{yr^{-1}}$) & ($M_{\odot}\,\mathrm{yr^{-1}}$) & ($M_{\odot}\,\mathrm{yr^{-1}}$) \\
    \hline
        REBELS-29$^*$\tnote{*} & $6.6847\pm0.0002$ & 150.403542 & 2.630306 & $10_{-3}^{+3}$ &  $35\pm3$ & $22.4^{+18.8}_{-10.5}$ & $61.0\pm7.0$ \\
        REBELS-29-2 & $6.6813\pm0.0005$ & 150.403875 & 2.632339 & $<25$ & $<2$ & $74.3^{+47.0}_{-31.2}$ & $100.0\pm11.2$ \\
        REBELS-12 & $7.347\pm0.001$ & 36.2830833 & -5.111316 & $19^{+9}_{-7}$ & $41\pm3$ & $70.5^{+51.6}_{-31.3}$ & $163.0\pm31.5$ \\
        REBELS-12-2 & $7.3521\pm0.0005$ & 36.2828333 & -5.114225 & $<10$ & $<14.7$ & $43.1^{+37.0}_{-22.7}$ & $88.0\pm16.6$ \\
    \hline
    \end{tabular}
    \label{tab:derivedproperties}
    \begin{tablenotes}
    \item{*} ID304384 in Bowler et al. (2018)\cite{Bowler2018}
    \item{$\dagger$} Spectroscopic redshift measured from [CII] $158\,\mathrm{\mu m}$ emission lines.
    \item{$\dagger\dagger$} Based on L$_\mathrm{IR}$ estimates with conservative errorbars (see Supplementary Material)
    \end{tablenotes}
\end{table*}

The serendipitous discovery of these two dusty galaxies at $z\sim7$ shows that our current (UV-based) census of very early galaxies is still incomplete. It is thus crucial to estimate the contribution of such sources to the total cosmic SFR density. However, this is not a trivial task given that these sources were only found as neighbors to UV-luminous primary targets. We therefore provide several different estimates (see Methods for details). 
Assuming that REBELS-29-2 and REBELS-12-2 were found in a completely blind survey consisting of all REBELS data-cubes and using their UV+IR-based SFRs, their SFRD would amount to $\sim$60-100\,\% of the total SFRD contributed from UV-selected LBGs, $\rho_{LBG}$\cite{Madau2014} (Figure~3). However, given that these sources were in fact detected as clustered galaxies in a targeted follow-up survey, we have to account for the excess probability of finding such sources based on the correlation function. Doing this, we estimate that the actual contribution by these sources to the SFRD is reduced by a factor $4.1\pm0.6$, i.e., amounting to $\sim10-25\%$ of $\rho_{LBG}$. 

An independent, conservative estimate of their SFRD contribution can be obtained from the fraction of primary UV-luminous targets with confirmed [CII] lines that revealed such a dusty neighbor (2 out of 19, i.e., $10.5^{+9.1}_{-5.0}\,\%$).  Assuming that this fraction also applies to fainter sources, this would result in an additional SFRD of 
$8.9^{+7.7}_{-4.3}\times10^{-4}\,\mathrm{M_{\odot}\,yr^{-1}\,Mpc^{-3}}$, i.e., $\sim11\%$ of UV luminous sources. All these estimates are starting to constrain different models for the contribution of dust-obscured sources at high redshifts\cite{Casey2018} (see Figure~3).  It is clear, however, that a blind, wide area survey for such sources is required in the future to properly constrain their number density in blank fields. These surveys must observe substantially deeper than had been envisioned previously\cite{Zavala21} to sample the fainter dust-obscured, but otherwise ``normal'' galaxies such as REBELS-12-2 and REBELS-29-2.

\begin{figure}[h!]
    \centering
    \includegraphics[width=1.0\columnwidth]{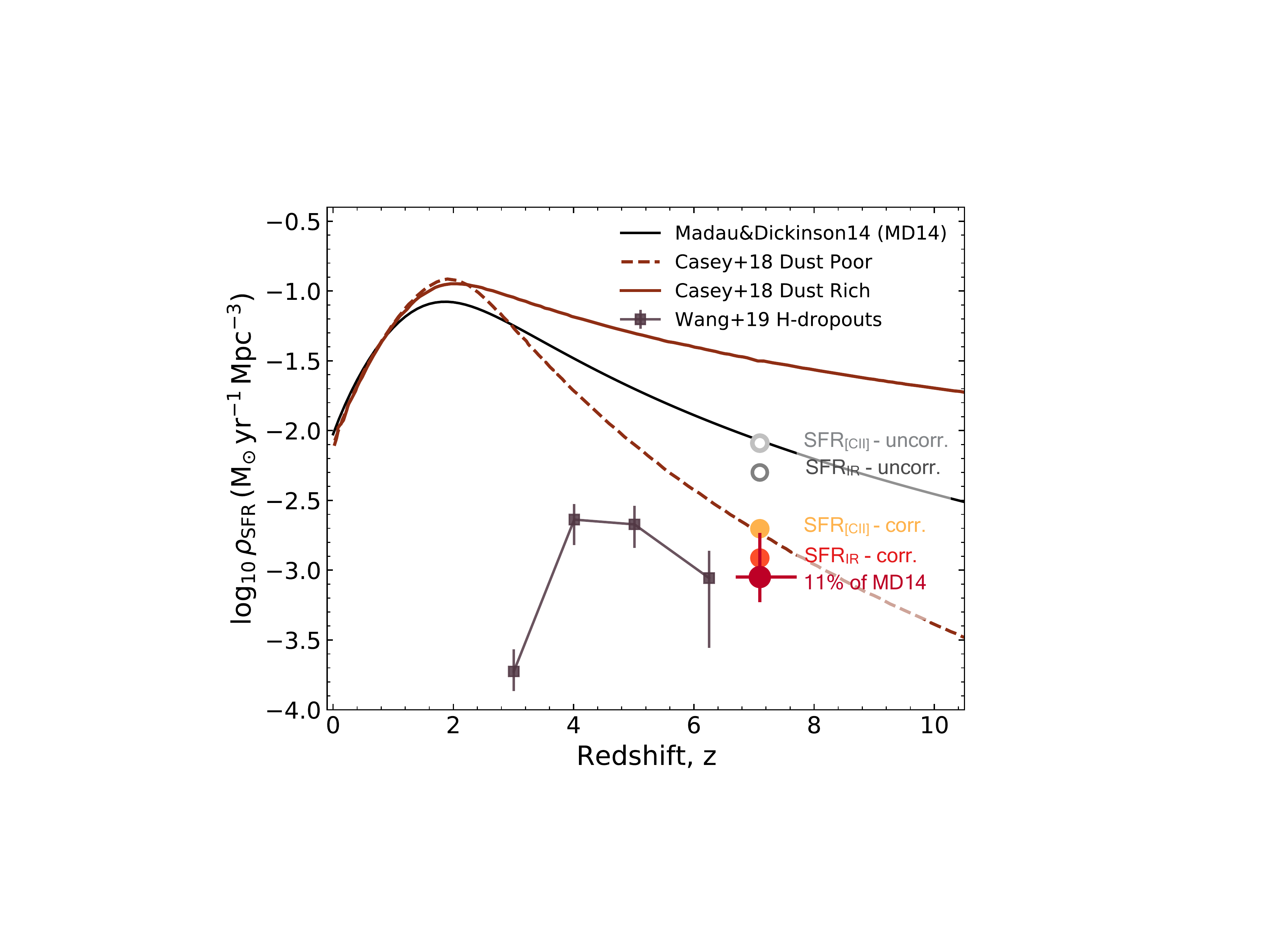}
    \caption{\small\textbf{Contribution of obscured galaxies to the cosmic SFR density $\mathbf{\rho_{\mathrm{SFR}}}$}: The black solid line shows the consensus estimate $\rho_{\mathrm{SFR}}^{\mathrm{MD14}}$ as a function of redshift\cite{Madau2014}, which at $z>4$ is derived from LBGs with a small dust correction. The dark red solid and dashed lines indicate two possible models for the extra contribution from obscured sources\cite{Casey2018}, which can be constrained from our observations. Error bars correspond to $1\sigma$ uncertainties. At most, the dusty $z\sim7$ galaxies identified here could contribute up to 60-100\% of $\rho_{\mathrm{SFR}}^{\mathrm{MD14}}$
depending on the exact estimate of their SFRs (dark gray, $\mathrm{SFR_{IR}}$, and light gray $\mathrm{SFR_{[CII]}}$). However, when correcting these values for the expected excess due to clustering, they lie a factor $\sim4\times$ lower (light red and orange dots). A similar SFRD estimate is found from the fraction of [CII]-confirmed REBELS targets that show a dusty companion (11\%; dark red). All these estimates are consistent with Spitzer-selected, massive dusty galaxies at $z\sim3-6$ (squares\cite{Wang2019}). While the exact SFRD contributed by dust-obscured sources is still uncertain, their existence in the epoch of reionization implies a revision of our understanding of early galaxy assembly.
    }
    \label{fig:SFRD}
\end{figure}

\vspace{3pt}
\noindent\rule{\linewidth}{0.4pt}
\vspace{3pt}



\putbib[MasterBiblio]


\end{bibunit}

\clearpage
\appendix

\begin{bibunit}

\noindent\textbf{\large Methods}

\section{Cosmology}
Throughout this paper we assume a concordance cosmology with $\Omega_{m}=0.3$, $\Omega_{\Lambda}=0.7$, $\mathrm{h}=0.7$.

\section{Target Selection}
REBELS-12 and REBELS-29 are among the sample selected in the ALMA large program REBELS (Reionization-Era Bright Emission Line Survey).
The entire sample of the REBELS program consists of 40 UV-luminous galaxies ($\gtrsim2\times L_{\ast}$) with photometric redshifts at $z\sim6.5-9$. The corresponding UV-derived SFRs are in excess of $\sim30\,\mathrm{M_{\odot}\,yr^{-1}}$, but they still correspond to ``normal" star-forming galaxies, i.e., they lie on the main-sequence of star-formation (see Figure~1).
An additional, important criterion for choosing the targets was a small uncertainty on their photometric redshifts to maximize the efficiency of the overall survey.
The REBELS program is still on-going, with $\sim85\%$ of the final data acquired so far. For this paper, we exploit the current dataset taken during cycle 7.

\section{Observations and Reduction of ALMA data}
Our ALMA observations were performed as part of the project 2019.1.01634.L during ALMA Cycle-7.
To scan the whole redshift probability of [CII] $158\,\mathrm{\mu m}$ emission lines, the receivers were tuned to cover $243.13-250.63\,\mathrm{GHz}$ for REBELS-29, and $217.91-238.29\,\mathrm{GHz}$ for REBELS-12. 
In the each frequency setup, the correlator was set to frequency domain mode with a band width of $1875\,\mathrm{MHz}$ in each spectral window, in order to sufficiently resolve [CII] emission lines.
The observations were carried out in ALMA configuration C43-2, with projected baselines ranging from $14\,\mathrm{m}$ to $313\,\mathrm{m}$.
With our observing frequency, flux calibration errors are $<10\,\%$.

The data were calibrated using the standard ALMA calibration pipeline implemented in the Common Astronomy Software Applications package (CASA)  version 5.6.1.
We imaged the continuum and data cubes using the CASA task \texttt{tclean} with natural weighting to maximize point source sensitivity.
The resulting synthesized beam sizes were $1.15^{\prime\prime}\times 1.38^{\prime\prime}$ and $1.25^{\prime\prime}\times 1.51^{\prime\prime}$ for REBELS-29 and REBELS-12, respectively.
During the imaging process, synthesized beam is deconvolved by applying a source detection threshold of three times background RMS of dirty images.
Because there are no extremely bright sources in our cubes and the side lobes of the synthesized beam contribute $<10\,\%$ of the sensitivity, the exact choice of the cleaning threshold does not affect our image products.

After subtracting the continuum using the CASA task \texttt{uvcontsub}, we extracted spectra in an iterative way to create robust apertures for the extraction.
From the extracted spectra, we created moment-0 map spanning over $\pm2\,\sigma$ velocity width of the emission line. In the next iteration, any other pixels having positive signals down to $2\,\sigma$ were added and the spectra were re-extracted.
This iteration continued until the integrated emission line flux converged. In all cases a few iterations were enough to achieve this.

\section{Optical and Near-Infrared Data}
Optical and near-infrared images are obtained from publicly available surveys. 
In particular, REBELS-29 lies in the 2 deg$^2$ COSMOS field covered by UltraVISTA\cite{McCracken12}, and REBELS-12 is in the VIDEO survey\cite{Jarvis13} within the XMM-Newton Large-Scale Structure (XMM-LSS) field 3. Both of these fields thus have relatively deep, ground-based NIR imaging data in the \textit{YJHK}$_s$ bands, from which the primary REBELS sources are selected. We use the DR4 images of the UltraVISTA survey. The 5$\sigma$ limiting magnitudes in the J-band are thus 26.0 and 24.9, respectively. 
REBELS-29 is covered by deep optical images with Megacam from the Canada-France-Hawaii Legacy Survey (CFHTLS\cite{Erben09}) and both sources have additional optical imaging from the Subaru Hyper Suprime-Cam (HSC) survey\cite{Aihara18}.
These fields have also been observed with Spitzer/IRAC from various programs over the past few years. In the VIDEO field only $3.6\,\mathrm{\mu m}$ and $4.5\,\mathrm{\mu m}$ data are available, while COSMOS has also been covered with IRAC at $5.8\,\mathrm{\mu m}$ and $8.0\,\mathrm{\mu m}$. In $3.6\,\mathrm{\mu m}$, the 5$\sigma$ limiting magnitudes measured in $2^{\prime\prime}.8$ diameter apertures are 25.5 and 24.3 mag, respectively.
For more details on the ground-based and Spitzer imaging over these fields see Bowler et al\cite{Bowler20} and Stefanon et al\cite{Stefanon19}.
Additionally, REBELS-29 is covered by HST observations in the F140W filter\cite{Bowler17}, reaching a 5$\sigma$ depth of 26.9 mag. We combine all the available data to constrain the panchromatic SEDs of both the primary REBELS targets as well as the serendipitous, dusty companions.

\section{Derivation of Physical Parameters}

\noindent\textbf{Star formation rates:}
SFRs are estimated based on the UV and FIR luminosities (including conservative assumptions about the IR SED shapes; see the section below and Extended Data Table 1).
Alternatively, we also estimate SFRs based on the correlation between SFR and L$_\mathrm{[CII]}$, previously measured at 4$<$z$<$6 \cite{Schaerer2020} and $z\sim0$ \cite{DeLooze2014}.
While this [CII] approach yields slightly higher values of $100\pm11\,\mathrm{M_{\odot}\,yr^{-1}}$ and $88\pm17\,\mathrm{M_{\odot}\,yr^{-1}}$, respectively, they are still consistent within uncertainties with those derived from the UV and IR continuum (Table 1).
Given the uncertain calibration of the L$_\mathrm{[CII]}$-SFR correlation of high-redshift galaxies, we use the SFRs directly derived from FIR luminosities as our fiducial SFRs.

\noindent\textbf{Stellar mass limits:}
To interpret the serendipitous dusty galaxies, we estimate upper limits for their stellar masses based on their optical, NIR, and FIR photometry and SED modeling. 
These limits are further tested with dynamical mass estimations.

We use the publicly available code Bayesian Analysis of Galaxies for Physical Inference and Parameter EStimation (\texttt{BAGPIPES}\cite{Carnall2018})
to derive SEDs consistent with all linear flux measurements and their uncertainties (see Extended Table 2), while the redshifts are fixed to the [CII] detections.
The input stellar population synthesis models for \texttt{BAGPIPES} are based on the 2016 version of the Bruzual \& Charlot library\cite{Bruzual03}, using a Kroupa\cite{Kroupa02} initial mass function.
We adopt constant SFH models with formation time as a free parameter, and we allow for  metallicities ranging from 0.1 to 2.5 $\times Z_\odot$.
Nebular continuum and line emission are added in a self-consistent manner\cite{Byler17} based on the photoionization code \texttt{CLOUDY}\cite{Ferland17} using the ionization parameter ($\log U$) as a free parameter.
Dust attenuation is included using the standard attenuation law for star-forming galaxies\cite{Calzetti00}, with the attenuation in the V-band ($A_V$) as a free parameter. We have also tested the impact of using a different dust attenuation model\cite{Charlot2000}, but obtained consistent mass limits.
Differential dust attenuation is allowed for stars in their birth clouds by a multiplicative factor ($\eta$).
Dust emission is then included self-consistently using a grid of SED models\cite{DrainLi07} under the assumption of energy balance, i.e., that the dust-absorbed energy is re-radiated in the far-infrared. The dust emission model has three parameters: the minimum intensity of starlight incident on the dust ($U_{min}$), the fraction of dust particles at this lowest intensity ($\gamma$), and the amount of PAH emission ($q_{PAH}$).  Overall, these SED fits thus have nine free parameters, for which we assumed very wide, uniform priors (see Extended Data Table 3 for input priors). 
With the current optical/NIR data we do not expect to constrain all these free parameters for the dust-obscured sources. However, our approach allows us to marginalize over these parameters in order to derive realistic upper limits of stellar masses.
Through this analysis, we find 90\% probability  upper limits of  $\mathrm{log} M_{\ast}/\mathrm{M_{\odot}} < 10.4 (10.0)$ for REBELS-29-2 (REBELS-12-2).
The results are listed in Table 1 in the main text and the posterior SEDs are shown in Figure~2 and Extended Data Figure~1.

Additionally, we derive dynamical masses to test the above stellar mass limits from SED analyses.
While dynamical masses are very uncertain given the current low resolution observations, they can provide a useful, independent check.
In particular, we estimated dynamical masses using the [CII] $158\,\mathrm{\mu m}$ emission velocity dispersion (Table 1) and [CII] emission sizes. [CII] emission sizes are measured using CASA task \texttt{uvmodelfit} assuming a 2D Gaussian with free parameters of total fluxes and full width at half maximum (FWHM) while centroids are fixed. REBELS-29-2 is unresolved and thus consistent with a point source in our resolution. To estimate the upper limit of dynamical masses of REBELS-29-2, we therefore assumed a synthesized beam FWHM as an upper limit size of REBELS-29-2. REBELS-12-2 is marginally resolved with the best fit Gaussian FWHM of $1.7^{\prime\prime}(\pm0.8^{\prime\prime})\times1.0^{\prime\prime}(\pm1.0^{\prime\prime})$ with a position angle of $6.3\pm40$ degree.  Following previous studies\cite{Wang2013,Capak15}, we used a simplified dyanmical mass estimate, namely: $M_{\mathrm{dyn}}=1.165\times10^5\,v_{\mathrm{circ}}^2\,D$ where $D$ is a [CII] emission diameter in $\mathrm{kpc}$, and the circular velocity $v_{circ}$ is approximated using velocity dispersion ($\sigma_{\mathrm{[CII]}}$) and inclination angle $i$ as $v_{\mathrm{circ}}=1.763\times\sigma_{\mathrm{[CII]}}/\mathrm{sin}(i)$. For the inclination angles, as the low resolution observations only provides unconstrained or uncertain values, we assumed a uniform distribution of $\mathrm{sin}(i)=0.45-1$ as estimated as an approximations from a dispersion dominated system to an edge on disk\cite{Capak15}. In this way, we found dynamical masses of $<1.1\times10^{11}\,\mathrm{M_{\odot}}$, and $9.4^{+10.9}_{-5.3}\times10^{10}\,\mathrm{M_{\odot}}$.
Assuming gas mass fraction of $60\,\%$\cite{Dessauges2020}, these dynamical masses provide stellar mass constraints of $<6.6\times10^{10}\,\mathrm{M_{\odot}}$, and $3.4^{+6.6}_{-3.1}\times10^{10}\,\mathrm{M_{\odot}}$ for REBELS-29-2 and REBELS-12-2, respectively. While still uncertain, these estimates are thus higher, but completely consistent with the stellar mass upper limits from the panchromatic SED analyses.

\noindent\textbf{Infrared Luminosity and Dust Masses:} Using the rest-frame $\sim158\,\mathrm{\mu m}$ continuum measurements we estimated total infrared ($\lambda=8-1000\,\mathrm{\mu m}$) luminosities following previous works\cite{Casey2012,Capak15}, which assume a range of gray body models consistent with galaxies observed at  $z<6$\cite{Schreiber18,Faisst2020}. In particular, we use SEDs for which the optical depth reaches unity at $\lambda_0=100\,\mathrm{\mu m}$.  
To construct FIR SED models, we assumed the following parameters: $\alpha$ the blue power-law, $\beta$ the long wavelength slope, and $T_{\mathrm{SED}}$ the luminosity weighted dust temperature.
We assumed a very conservative range of these parameters uniformly distributed between $\alpha=1.5-2.5$, $\beta=1.5-2.5$, and $T_{\mathrm{SED}}=35-73\,\mathrm{K}$.
The range of luminosity weighted dust temperatures $T_{\mathrm{SED}}$  used here corresponds to a range of ``peak temperatures'' of $T_\mathrm{peak} \simeq 30-50\,\mathrm{K}$, as calculated from the peak wavelength of the SED and assuming Wien's displacement law ($T_\mathrm{peak}\, [K] = 2.9\times10^3 / \lambda_\mathrm{peak}\,[\mu m]$).
Using the assumed parameter distributions, we normalized the FIR SEDs to the ALMA continuum fluxes at the observed wavelength of $\lambda_{\mathrm{obs}}=1248\,\mathrm{\mu m}$ and $\lambda_{\mathrm{obs}}=1315\,\mathrm{\mu m}$ for galaxies in the REBELS-29 and REBELS-12 fields, respectively, and calculated IR luminosity distributions. We derived IR luminosities from the obtained IR luminosity distributions by calculating the posterior median, 16th percentile, and 84th percentile values.
In both the IR luminosity and dust mass calculations (see next paragraph), we applied corrections due to the CMB heating and CMB background against which we observe the dust continuum\cite{Dacunha2013}. The results are listed in Extended Data Table 1.

We derived dust masses using the same distribution of SED parameters assuming a dust mass absorption coefficient of $\kappa_{d}(\lambda_{\mathrm{rest}})=0.77\times((1+z) \times850/\lambda_{\mathrm{obs}})^{\beta}\,\mathrm{cm^{2}\,g^{-1}}$ at the observed wavelength of $\lambda_{\mathrm{obs}}=1248\,\mathrm{\mu m}$ and $\lambda_{\mathrm{obs}}=1315\,\mathrm{\mu m}$ for galaxies in REBELS-29 and REBELS-12 fields, respectively.
In particular, we assumed that the mass weighted dust temperatures have the same conservative distribution as the luminosity weighted dust temperatures.
The estimated dust masses are $M_{\mathrm{dust}}=2.2^{+2.2}_{-1.1}\times10^7$ and $1.2^{+1.7}_{-0.8}\times10^7\,\mathrm{M_{\odot}}$, consistent with previous studies of sources at similar redshift\cite{Watson15,Laporte17}. 
Note that these masses are likely lower limits of the total dust budget, as diffuse and relatively cold dust components would be in thermal equilibrium with the CMB \cite{Behrens2018,Liang2019,Sommovigo2020} and thus invisible from observations. Nevertheless, even with the current dust mass estimates, $>16-50\,\%$ of metals ever produced by supernovae in these sources are already locked into dust grains, consistent with local Universe values\cite{deVis19}. 
This suggests very fast dust build-up at $z>6$\cite{Mancini16,Graziani20}, which will be further investigated in a follow-up paper.

\section{Possible Lower Redshift Contamination?}
In the main text, we argued that the two serendipitously detected sources lie at the same redshift as the primary REBELS targets. The main reason for this is that galaxies in the Universe are clustered and that the emission lines lie at almost exactly the same frequency (within less than 0.2GHz) of the main targets' lines, corresponding to velocity offsets of $<$250\,km\,s$^{-1}$. 
Here, we further quantify the probability of detecting a lower redshift source that has an emission line within 0.2 GHz of the primary target's [CII] frequency and within the HPBW of our ALMA observations ($\sim13$ arcsec radius). 
We base this analysis on the ALMA continuum detections of the two sources, and the IR luminosity functions measured at lower redshifts\cite{Gruppioni13}.
The most likely candidate for random emission lines are transitions of CO. However, we also test other lines that have been detected in distant galaxies\cite{Carilli13}. For each possible line, we compute the corresponding source redshift and then convert the ALMA continuum emission to a total infrared luminosity using the same template SED. Based on the IR LFs, we then compute the expected number of galaxies with ALMA fluxes larger than the observed source within a redshift interval such that the emission line would lie within 0.2 GHz of the main target's line. As an example, if the emission line of REBELS-29-2 were CO(8-7), the source would lie at $z=2.73$ and have an infrared luminosity of $\log L_{IR}/L_\odot=12.0$. Based on the $z\sim3$ IR LF, one would then expect 3.3$\times10^{-4}$ galaxies per random ALMA pointing with a continuum flux larger than what is observed.
For all the lines we tested, these numbers turn out to be similarly small ($<6\times10^{-4}$ galaxies) for both REBELS-12-2 and REBELS-29-2. Expressed in another way, one would need to observe $>1600$ ALMA pointings to find \emph{one} galaxy with a continuum flux density higher than either of the two serendipitous sources and with an emission line within 0.2 GHz of the primary targets. Hence, we can safely exclude the possibility that these lines stem from random foreground galaxies.

\section{A New Parameter Space of Dusty Galaxies}
As discussed in the main text, the two sources REBELS-29-2 and REBELS-12-2 are likely higher redshift analogues of the dust-obscured galaxies previously identified based on photometric redshifts at $z\sim3-6$\cite{Wang2019}. In Extended Data Figure~3, we compare these sources to different galaxy samples from the literature. 
In particular, we also show three of the four dusty companion galaxies to $z=6.0-6.6$ QSOs reported in Decarli et al.\cite{Decarli2017} that remained undetected in follow-up rest-frame UV observations\cite{Mazzucchelli2019}. Those sources are likely much rarer than our galaxies detected here. The number density of their central QSOs is two orders of magnitude lower than the UV-luminous LBGs of our sample.
Additionally, the newly found dusty companions are located at the lower end of QSO companions in terms of infrared and [CII] luminosities.
Therefore, as can be seen in Extended Data Figure~3, our sources at $z\sim7$ probe a different parameter space in terms of stellar mass and IR luminosities than previous samples.

Even though uncertain, the SEDs of our sources are consistent with being lower luminosity versions of the typical dust-obscured, starburst galaxies at $z\gtrsim6$ (Figure~2).
Given the current depth of the rest-frame UV images, $>97\,\%$ and $>75\,\%$ of their star formation activities are obscured ($95\,\%$ confidence lower limit).
This is in stark contrast to the UV-luminous targets, REBELS-12 and REBELS-29, for which only $39\,\%$ to $63\,\%$ of star formation is obscured (see Extended Data Figure~4). 
This means that $z\gtrsim6$ DSFGs need to be searched for in deeper surveys than previously assumed. The main question now is: How common are such lower-luminosity DSFGs in the early Universe? In the next section, we provide several estimates of their contribution to the cosmic SFRD. 

\section{Contribution to the Cosmic SFR Density}
The calculation of the cosmic SFRD contributed by the dusty REBELS galaxies is not trivial. In particular, because these galaxies were only discovered as neighbors to UV-luminous primary sources in our targeted follow-up program. As discussed in the main text, we derive several estimates, which are detailed below.

A somewhat naive, first estimate can be obtained, if we assume that these two sources were detected at random in a blind survey. In this case, the SFRD is simply given by their summed SFR divided by the whole survey volume spanned by the current REBELS dataset. For the survey volume, we integrate over the full frequency ranges over which we scanned for [CII] $158\,\mathrm{\mu m}$ emission lines and we use the area covered by the half primary beam widths (HPBWs). This totals to a survey volume of $2.3\times10^{4}\,\mathrm{Mpc^3}$. Then, summing the UV+IR-based SFRs of REBELS-29-2 and REBELS-12-2, we obtain $\rho_\mathrm{SFR}=5.1\times10^{-3}\,\mathrm{M_{\odot}\,yr^{-1}\,Mpc^{-3}}$. Using the [CII]-based SFRs, this amounts to $\rho_\mathrm{SFR}=7.8\times10^{-3}\,\mathrm{M_{\odot}\,yr^{-1}\,Mpc^{-3}}$. However, as indicated above, these numbers should be taken as upper limits given that REBELS is not a blind survey.

To account for clustering, we can obtain an estimate for the expected boost in number counts (and hence the SFRD) compared to a random, blind survey based on the correlation function\cite{Peebles80}. Starting from the real-space correlation function $\xi(r)=(r/r_0)^{-\gamma}$ as measured for $z\sim7$ galaxies\cite{BaroneNugent14}, we perform the Limber transform to derive the corresponding angular correlation function\cite{Adelberger05}. This requires a redshift selection function, which we assume to be a top-hat with redshift depth of $\Delta z=0.33$, i.e., the minimum frequency coverage of the REBELS [CII] line search. 
The expected number of neighbors in excess of a random field is then derived from this angular correlation function by integrating over a solid angle with radius corresponding to the HPBW (13\arcsec). Using the measured correlation function parameters $r_0=6.7\pm0.9\,h^{-1}$cMpc and $\gamma=1.6$, this results in a boost factor due to clustering of 4.1$\pm$0.6$\times$. Hence, the SFRD estimates above need to be corrected down by this amount, resulting in $\rho_\mathrm{SFR}=1.2\times10^{-3}\,\mathrm{M_{\odot}\,yr^{-1}\,Mpc^{-3}}$ or $\rho_\mathrm{SFR}=1.9\times10^{-3}\,\mathrm{M_{\odot}\,yr^{-1}\,Mpc^{-3}}$, respectively, using the UV+IR based SFRs or the [CII]-derived SFRs.

Unfortunately, the correction factor due to clustering depends quite sensitively on the assumed correlation function parameters, which themselves depend on the mass and luminosity of the sources\cite{Qiu18}. For instance, using the simulated correlation function\cite{Bhowmick20} for UV-luminous galaxies with $M_\mathrm{UV}<-22$ (corresponding to SFR$_{UV}>24$M$_\odot$yr$^{-1}$), as appropriate for the REBELS primary sample, we derive a boost factor as high as 30$\times$. However, we note that these simulations do not include such dusty sources (which have $M_\mathrm{UV}>-19.4$ and $M_\mathrm{UV}>-21.4$, respectively), and we will defer the reader to a later paper to estimate an appropriate correlation function for DSFGs such as REBELS-29-2 and REBELS-12-2.

Another, more conservative SFRD estimate of dusty galaxies can be obtained from the fraction of REBELS data cubes that showed such sources, and assuming that these are representative of the cosmic average. Given the existing data from our on-going program, we currently have [CII] emission lines confirmed  in 19 primary targets (at $>6\sigma$). We performed a blind search for other  lines in the current data set, but have only found REBELS-12-2 and REBELS-29-2 without optical counterparts. Hence, for 19 UV-luminous galaxies we found 2 dust-obscured counterparts with similar masses and SFRs. If we extrapolate this to the full LBG population, this would imply that such dusty sources contribute $10.5_{-5.0}^{+9.1}\,\%$ of $\rho_\mathrm{SFR}^{MD14}$, i.e. 
 $8.9^{+7.7}_{-4.3}\times10^{-4}\,\mathrm{M_{\odot}\,yr^{-1}\,Mpc^{-3}}$. Note that within the large uncertainties, this value is completely consistent with our clustering-corrected SFRDs derived above.

As a final reference, we can also derive an upper limit from the ALMA large program ASPECS, which performed a completely blind line scan over the HUDF, covering [CII] in the redshift range $z=6-8$. This scan would have been sensitive to galaxies with [CII]-based SFRs $\gtrsim16\mathrm{M}_{\odot}\,\mathrm{yr}^{-1}$ (5$\sigma$ detection limit). However, no such sources were found\cite{Uzgil21}.
Using the full ASPECS survey volume, one can thus derive a limit on the total cosmic SFRD from ASPECS down to these SFRs, which results in $<1.0\times10^{-2}\,\mathrm{M_{\odot}\,yr^{-1}\,Mpc^{-3}}$. 

All the estimates derived above, including the ASPECS blind search, lie significantly below the ``Dust-Rich" model from Casey et al.\cite{Casey2018}, which assumes that dust-obscured galaxies contribute $\sim90\,\%$ of the total SFR density at $z>4$. On the other hand, our clustering-corrected estimates are in good agreement with the ``Dust-Poor" model.
However, we note that our sample is limited in SFR and that we needed to extrapolate to lower luminosity sources, as outlined above.
At lower redshifts, the fraction of obscured star-formation decreases very rapidly, however, to lower mass and luminosity systems\cite{Whitaker2017,Fudamoto2020}. Based on this, it would in principle not be expected that the integration to fainter systems would increase the SFRD of dust-obscured galaxies. However, as we show in (see Extended Data Figure~4), the obscured fraction of star-formation may indeed show a much larger variation than previously expected, given that earlier estimates were mostly based on UV-selected samples. Clearly, larger datasets are required to test this further in the future.

\vspace{3pt}
\noindent\rule{\linewidth}{0.4pt}
\vspace{3pt}
\let\oldthebibliography=\thebibliography
\let\oldendthebibliography=\endthebibliography
\renewenvironment{thebibliography}[1]{%
    \oldthebibliography{#1}%
    \setcounter{enumiv}{23} 
}{\oldendthebibliography}
\putbib[only_in_appendix]
%
\begin{addendum}
 \item[Acknowledgements] The authors thank Christina Williams for helpful discussions.  YF and PAO acknowledge support from the Swiss National Science Foundation through the SNSF Professorship grant 190079 `Galaxy Build-up at Cosmic Dawn'. YF further acknowledges support from NAOJ ALMA Scientific Research Grant number 2020-16B `ALMA HzFINEST: High-z Far-Infrared Nebular Emission STudies'.
 PD, AH and GU acknowledge support from the European Research Council's starting grant ERC StG-717001 (DELPHI). PD also acknowledges support from the NWO's VIDI grant (``ODIN"; 016.vidi.189.162) and the European Commission's and University of Groningen's CO-FUND Rosalind Franklin program.
 LG and RS acknowledge support from the Amaldi Research Center funded by the MIUR program "Dipartimento di Eccellenza" (CUP:B81I18001170001).
 HI acknowledges support from JSPS KAKENHI Grant Number JP19K23462.
 RS acknowledges support from an STFC Ernest Rutherford Fellowship (ST/S004831/1).
 M.A. has been supported by the grant “CONICYT + PCI + INSTITUTO MAX PLANCK DE ASTRONOMIA MPG190030” and “CONICYT+PCI+REDES 190194.”.
The Cosmic Dawn Center (DAWN) is funded by the Danish National Research Foundation under grant No.\ 140.

 \item[Author Information] The authors declare that they have no competing financial interests. Correspondence and requests for materials should be addressed to Y.F.~(email: y.fudamoto@aoni.waseda.jp).

\item[Author Contributions] Y.F. wrote the main part of the text, analyzed the data, produced most of the figures.  P.A.O. contributed text and led the SED fitting and data analysis. S.S. calibrated the ALMA data and produced images. M.S. performed detailed photometric measurements from the ground-based images. R.S. contributed comparison plots of different galaxy samples.  All co-authors contributed to the successful execution of the ALMA program, to the scientific interpretation of the results, and helped to write up this manuscript.

 \item[Data availability] The datasets generated during and/or analysed during the current study are available from the corresponding author on reasonable request. This paper makes use of the following ALMA data: ADS/JAO.ALMA \#2019.1.01634.L. ALMA is a partnership of ESO (representing its member states), NSF (USA) and NINS (Japan), together with NRC (Canada), MOST and ASIAA (Taiwan), and KASI (Republic of Korea), in cooperation with the Republic of Chile. The Joint ALMA Observatory is operated by ESO, AUI/NRAO and NAOJ.
 
 \item[Code availability] The codes used to reduce and analyse the ALMA data are publicly available. The code used to model the optical-to-infrared SEDs is accessible through github (\url{https://github.com/ACCarnall/bagpipes}).
\end{addendum}

\newpage
\onecolumn
\noindent\textbf{\large Extended Data}
\renewcommand\thefigure{\arabic{figure}}
\setcounter{figure}{0}
\renewcommand{\figurename}{Extended Data Figure}

\begin{figure*}[h]
    \centering
    \includegraphics[width=0.65\textwidth]{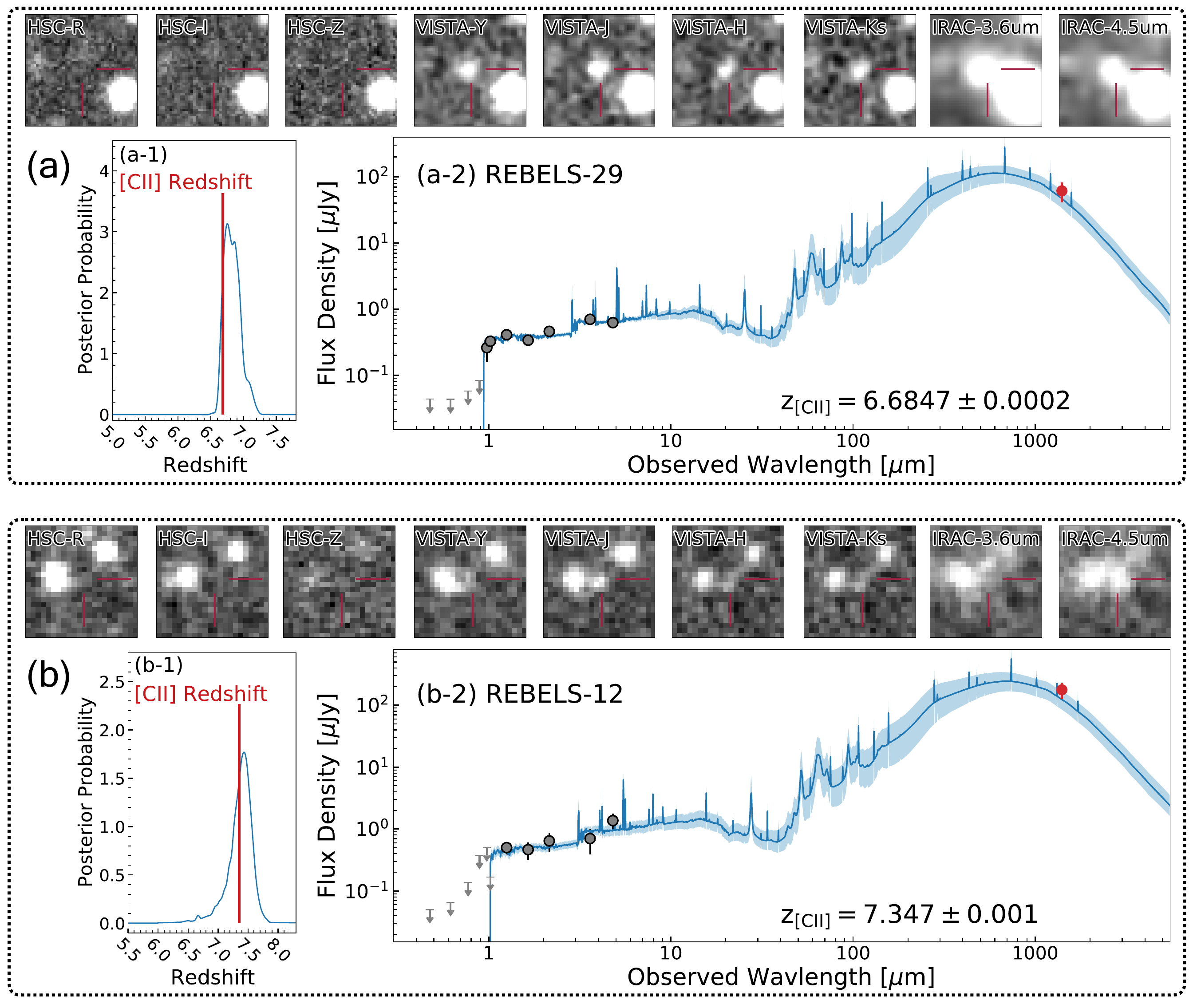}
    \caption{\small\textbf{Optical/NIR images and full SEDs of the UV-luminous targets REBELS-29 and REBELS-12.} 
The cutouts show images from which photometry was extracted. SED fits (bottom right panels) are performed using the BAGPIPES\cite{Carnall2018}.
In {\bf b} and {\bf d}, blue solid lines and bands represent the median posterior SEDs together with their $68\,\%$ confidence contours for REBELS-29 and REBELS-12, respectively. Error bars corresponds to $1\sigma$ uncertainties, and downward arrows show $2\sigma$ upper limits. {\bf a} and {\bf c} show that the [CII]$158\,\mathrm{\mu m}$ emission line redshifts (red) are in perfect agreement with the photometric redshift probability distributions (blue), that had been previously estimated from the optical/NIR photometry for both sources. This confirms their high-redshift nature.}
    \label{fig:cutouts_UV}
\end{figure*}
\clearpage
\begin{figure*}[t]
    \centering
    \includegraphics[width=0.7\textwidth]{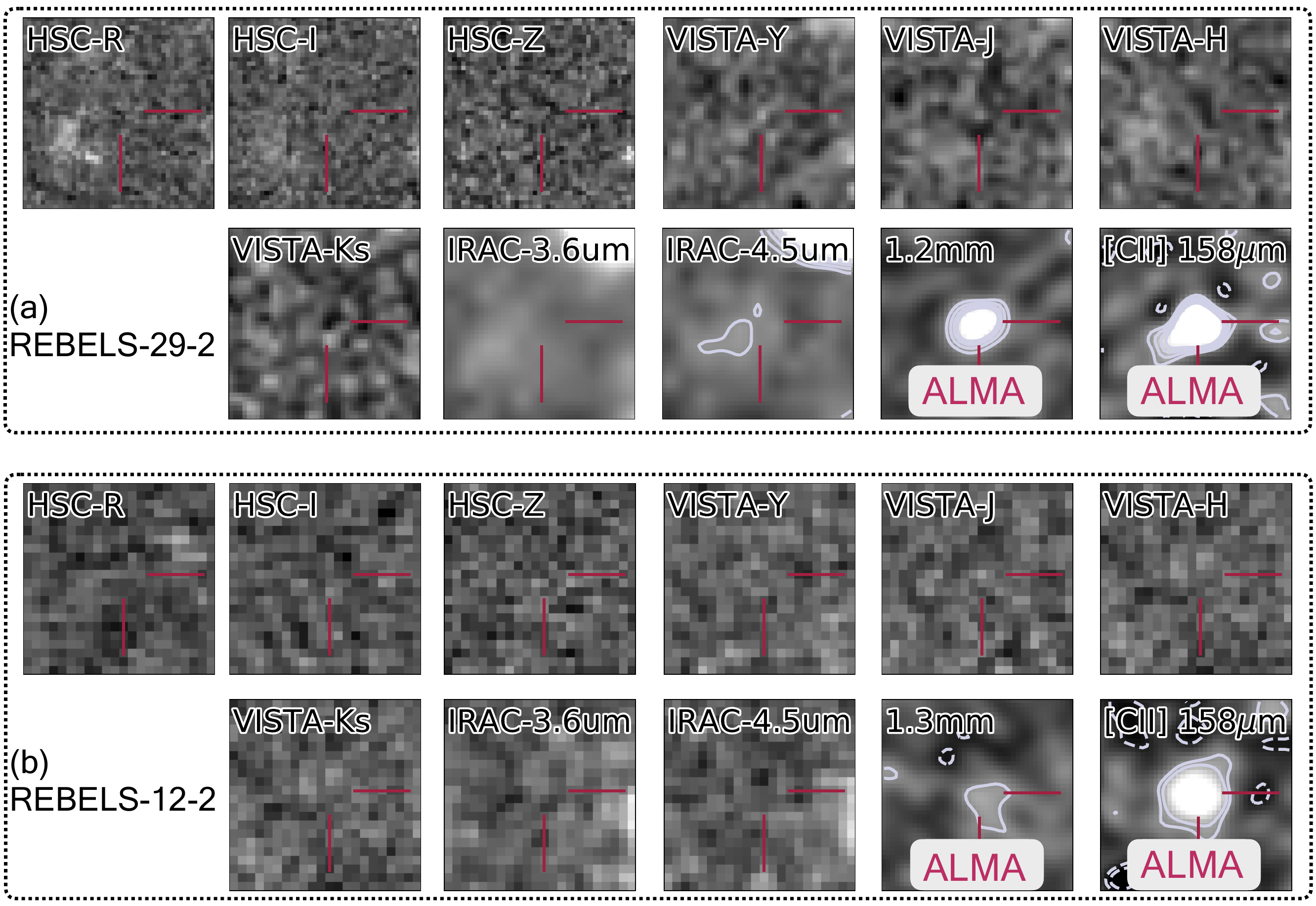}
    \caption{
    \small\textbf{Optical/NIR/FIR cutouts of the dusty sources REBELS-29-2 and REBELS-12-2.} $6.5^{\prime\prime}\times6.5^{\prime\prime}$ cutouts show the existing ground- and space-based observations: Subaru Hyper Suprime Cam, VISTA VIRCAM, Spitzer IRAC, in addition to the ALMA dust continuum images and continuum subtracted [CII] $158\,\mathrm{\mu m}$ moment-0 images.
White contours show $+2,+3,+4,+5\,\sigma$ (solid contour) and $-5,-4,-3,-2\,\sigma$ (dashed contour), if present. A faint low-surface brightness foreground neighbor can be seen $\sim2.0^{\prime\prime}$ to the SE of REBELS-29-2.
However, the photometric redshift of this foreground source is $z_{\mathrm{ph}}=2.46^{+0.08}_{-0.07}$, and the line frequency of REBELS-29-2 is not consistent with bright FIR emission lines (e.g. CO lines) from this foreground redshift.
No optical counterparts are found at the location of the ALMA [CII] and dust continuum positions for both REBELS-29-2 and REBELS-12-2. }
    \label{fig:cutouts_dusty}
\end{figure*}

\clearpage
\begin{figure*}[h]
    \centering
    \includegraphics[width=0.85\textwidth]{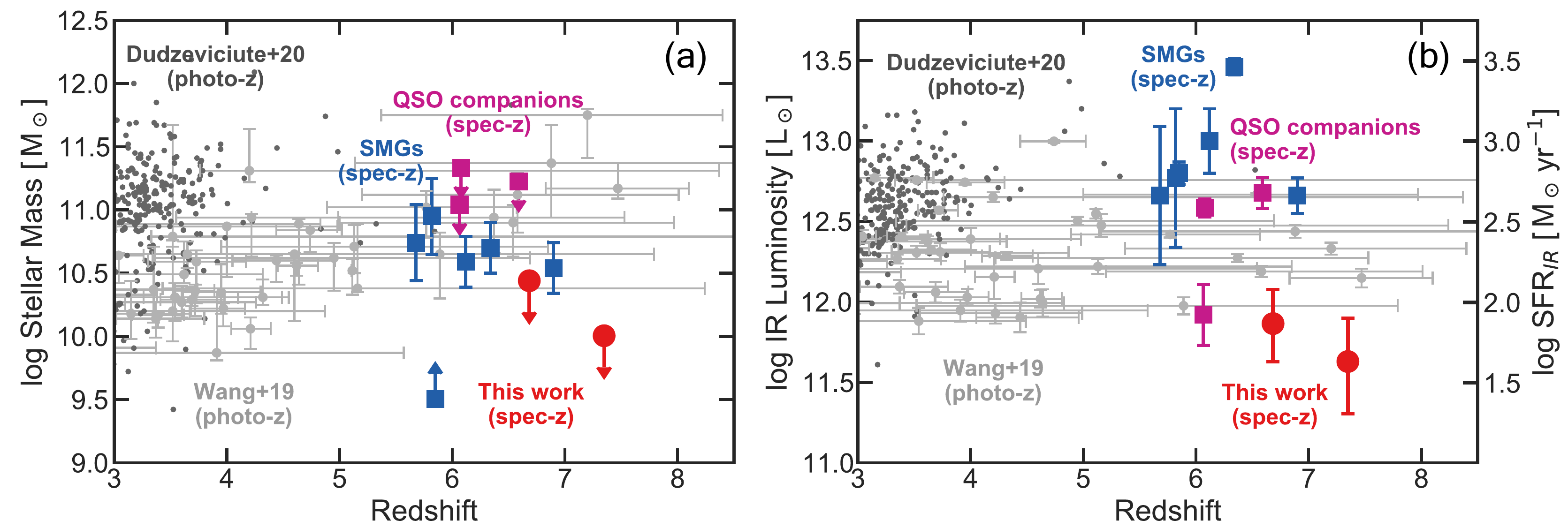}
    \caption{\small \textbf{Probing a new parameter space of DSFGs.} {\bf a:} The stellar mass as a function of redshift for DSFGs from the literature. IRAC-selected, H-dropout galaxies (light gray dots with 1$\sigma$ errorbars\cite{Wang2019}) are generally more massive than the two serendipitously detected REBELS galaxies (red dots). Additionally, the redshifts of H-dropouts are extremely uncertain (photo-z). The extremely star-bursting SMG population only shows a small tail of rare sources at $z>4$ (shown by dark dots\cite{Dudzeviciute20}). The blue squares show all the previously known DSFGs at $z>5.5$ with spectroscopically measured redshifts, while purple squares correspond to $z\sim6$ QSO companion galaxies\cite{Decarli2017}. These are more extreme sources than REBELS-12-2 and REBELS-29-2.
{\bf b:} The infrared luminosity / SFR$_\mathrm{IR}$ as a function of redshift for the same galaxy samples as on the left.  The infrared luminosities and hence SFRs of the newly identified galaxies are significantly lower than typical SMGs at these redshifts.
For both panels, error bars correspond to $1\sigma$ uncertainties, and arrows show $2\sigma$ upper/lower limits.
    }
    \label{fig:DSFGcomparison}
\end{figure*}
\vspace{2cm}
\begin{figure*}[h!]
    \centering
    \includegraphics[width=0.4\textwidth]{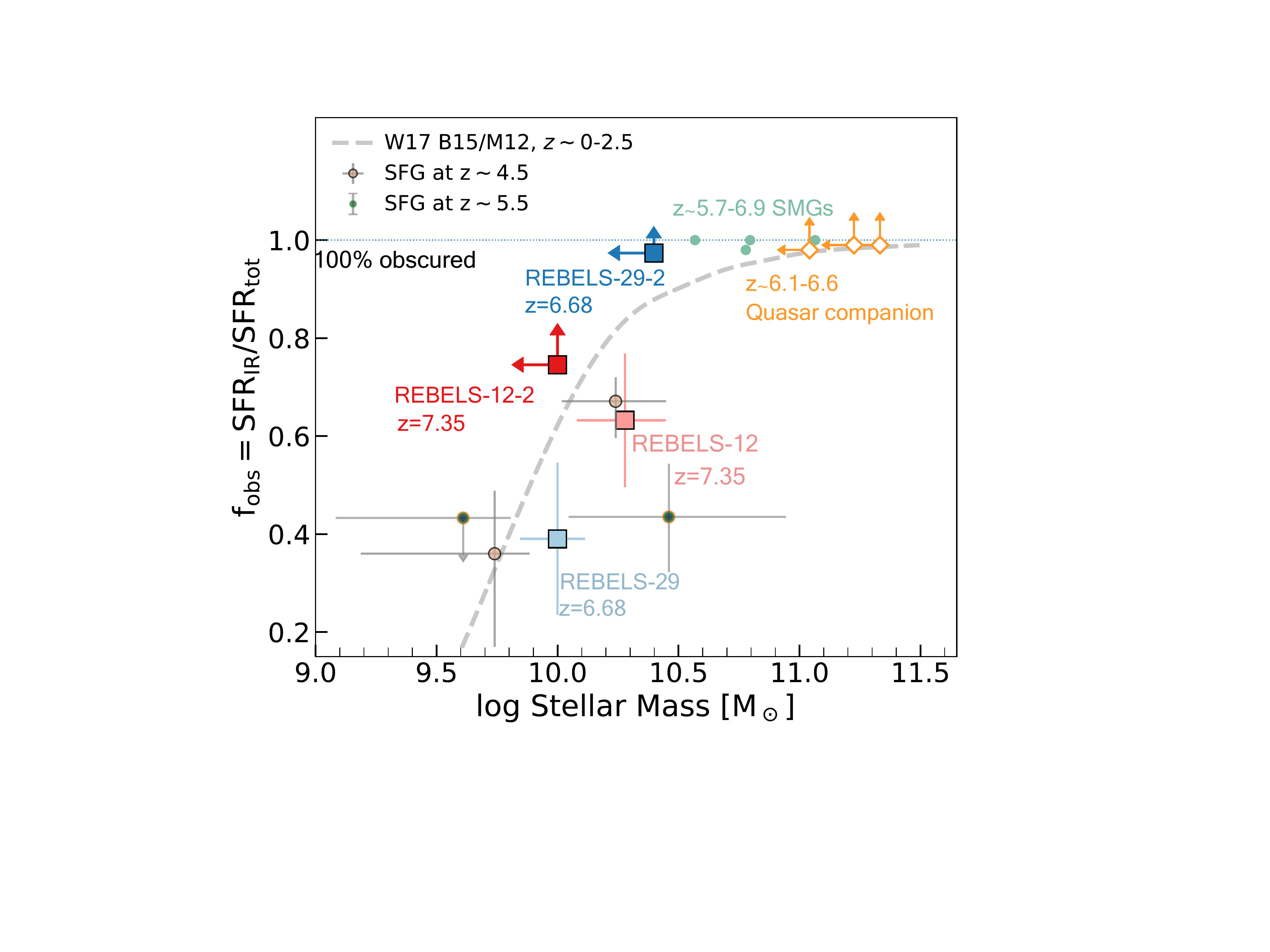}
    \caption{\small \textbf{Fraction of obscured star-formation as a function of stellar mass.}
The fraction of obscured star-formation, $\mathbf{f_{\mathrm{obs}}}=\mathrm{SFR_{\mathrm{IR}}}/(\mathrm{SFR_{\mathrm{IR}}} + \mathrm{SFR_{\mathrm{UV}}})$, of REBELS-29-2 and REBELS-12-2 (dark colored squares) is significantly higher than for typical LBGs at their stellar mass.
The line shows the observed, constant relation between $z\sim0$ and $z\sim2.5$\cite{Whitaker2017} assuming a given set of SED templates from Bethermin et al.\cite{Bethermin15}.
Blue and brown small points with error bars show stacked results of star forming galaxies at $z\sim4.5$ and at $z\sim5.5$, respectively \cite{Fudamoto2020}.
The star-formation of extreme starburst galaxies at $z\sim5.7-6.9$ is essentially 100\% obscured (SMGs\cite{Riechers2020}; green small points). 
The highly obscured star forming galaxies found as companions of high-redshift quasars at $z>6$\cite{Decarli2017,Mazzucchelli2019} (yellow diamonds) are significantly more massive than the galaxies identified here, as estimated from their dynamical masses.
Squares show the obscured fraction of our UV-bright and dusty galaxies.
Error bars correspond to $1\sigma$ uncertainty, and arrows show $2\sigma$ lower/upper limits.
Our discovery of lower mass, obscured galaxies shows that $f_{\mathrm{obs}}$ likely varies much more strongly at a fixed stellar mass than previously estimated even in the epoch of reionization.}
    \label{fig:fobs}
\end{figure*}
\clearpage
\begin{table*}[h]
    \centering
    \footnotesize
    \begin{tabular}{lclllccc}
    \multicolumn{8}{l}{\textbf{Extended Data Table 1: FIR Observed Properties (ALMA measurements)}} \\[0.1cm]
    \hline
        Galaxy Name & $\nu_{\mathrm{obs}}$$^{\dagger}$ & $\sigma_{\mathrm{[CII]}}$$^{\dagger}$ & $f_{\mathrm{[CII]}}$ & $f_{\mathrm{cont}}$ & $L_{\mathrm{[CII]}}$ & $L_{\mathrm{IR}}$ 
        &Dust Mass$^*$\tnote{*}\\
         & (GHz) & (km/s) &  ($\mathrm{Jy\,km/s}$) & ($\mathrm{\mu Jy}$) & ($10^{8}\,\mathrm{L_{\odot}}$) & ($10^{11}\,\mathrm{L_{\odot}}$) 
         & ($10^{7}\,\mathrm{M_{\odot}}$)\\
    \hline
        REBELS-29 & $247.30\pm0.01$ & $63\pm9.5$ & $0.44\pm0.06$ & $61\pm20$ & $4.86\pm0.64$ & $2.21^{+1.85}_{-1.01}$ & $0.7^{+0.7}_{-0.4}$\\
        REBELS-29-2 & $247.39\pm0.02$ & $142\pm18$ & $0.78\pm0.10$ & $192\pm25$ & $8.56\pm1.13$ & $7.33^{+4.63}_{-3.08}$ & $2.2^{+2.2}_{-1.1}$ \\
        REBELS-12 & $227.72\pm0.03$ & $359\pm45$ & $1.20\pm0.27$ & $177\pm52$ & $15.1\pm3.40$ & $6.87^{+4.91}_{-2.95}$ & $2.1^{+2.3}_{-1.1}$\\
        REBELS-12-2 & $227.56\pm0.02$ & $130\pm21$ & $0.58\pm0.13$ & $110\pm52$ & $7.33\pm1.60$ & $ 4.26^{+3.65}_{-2.24}$ & $1.2^{+1.7}_{-0.8}$\\
    \hline
     \multicolumn{8}{l}{All values are corrected for the primary beam attenuation and CMB, with $T_d=40\,\mathrm{K}$ and $\beta=1.5$,} if necessary.\\
     \multicolumn{8}{l}{{$\dagger$} $\nu_{\mathrm{obs}}$ and $\sigma_{\mathrm{[CII]}}$ are measured by Gaussian fitting the extracted spectra.}
    \end{tabular}
    \label{tab:derivedproperties2}
\end{table*}
\vspace{3cm}
\begin{table*}[h]
    \centering
    \footnotesize
    \begin{tabular}{lcrrrrl}
    \multicolumn{5}{l}{\textbf{Extended Data Table 2: Near Infrared Photometric Data}} \\[0.1cm]
    \hline
        Band & Wavelength [\AA] & REBELS-29$^{\dagger}$ & REBELS-29-2$^{\dagger}$ & REBELS-12$^{\dagger\dagger}$ & REBELS-12-2$^{\dagger\dagger}$ & Telescope\\
    \hline
        g-band & 4816.12 & $0.01 \pm 0.02$ & $0.02 \pm 0.02$ & $-0.01 \pm 0.02$ & $-0.03 \pm 0.03$ & Subaru/HSC\\
        r-band & 6234.11 & $0.01 \pm 0.02$ & $0.00 \pm 0.02$ & $0.01 \pm 0.03$ & $0.00 \pm 0.03$ & Subaru/HSC\\
        i-band & 7740.58 & $0.02 \pm 0.03$ & $0.01 \pm 0.03$ & $-0.05 \pm 0.07$ & $-0.01 \pm 0.07$ & Subaru/HSC\\
        z-band &9125.20 & $0.00 \pm 0.04$ & $-0.04 \pm 0.04$ & $0.01 \pm 0.19$ & $0.06 \pm 0.19$ & Subaru/HSC\\
        Y-band & 10214.19 & $0.33 \pm 0.03$ & $0.05 \pm 0.03$& $0.14 \pm 0.08$ & $0.04 \pm 0.09$ & VISTA/VIRCAM\\
        J-band & 12534.65 & $0.41 \pm 0.04$ & $-0.03 \pm 0.04$ & $0.50 \pm 0.11$ & $0.08 \pm 0.12$ & VISTA/VIRCAM\\
        H-band & 16453.41& $0.34 \pm 0.05$ & $-0.08 \pm 0.05$ & $0.46 \pm 0.15$ & $0.10 \pm 0.19$ & VISTA/VIRCAM\\
        Ks-band &21539.88 & $0.46 \pm 0.06$ & $-0.06 \pm 0.06$ & $0.64 \pm 0.22$ & $-0.29 \pm 0.23$ & VISTA/VIRCAM\\
        $3.6\,\mathrm{\mu m}$ & 35634.28 & $0.70 \pm 0.04$ & $0.08 \pm 0.05$ & $0.70 \pm 0.31$ & $0.13 \pm 0.29$ & Spitzer/IRAC\\
        $4.5\,\mathrm{\mu m}$ & 45110.13 & $0.62 \pm 0.06$ & $0.14 \pm 0.07$ & $1.37 \pm 0.41$ & $0.34 \pm 0.36$ & Spitzer/IRAC\\
    \hline
    \multicolumn{6}{l}{Flux densities are shown in units of $\mathrm{\mu Jy}$}\\
    \multicolumn{6}{l}{$\dagger$ Images are available as a part of the COSMOS survey\cite{Scoville07a} and UltraVISTA DR4\cite{McCracken12}}\\
    \multicolumn{6}{l}{$\dagger\dagger$ Images are available as a part of the VIDEO survey\cite{Jarvis13}}.
    \end{tabular}
    \label{tab:photometricdata}
\end{table*}
\clearpage
\begin{table*}[t]
    \centering
    \footnotesize
    \begin{tabular}{lll}
    \multicolumn{3}{l}{\textbf{Extended Data Table 3: Priors used for panchromatic SED modeling}} \\[0.1cm]
    \hline\hline
    Assumed Physical Properties &&\\
    \hline
    Initial mass function & \multicolumn{2}{r}{Kroupa \& Boily\cite{Kroupa02}}\\
    Stellar population synthesis model & \multicolumn{2}{r}{Bruzual \& Charlot\cite{Bruzual03}}\\
    Dust attenuation law& \multicolumn{2}{r}{Calzetti et al.\cite{Calzetti00}}\\
    Star formation history & \multicolumn{2}{r}{Constant star formation}\\
    \hline\hline
    Parameters & Priors & Description\\
    \hline
    Minimum Age & [0.1,1.0] & Time since star formation switched on in Gyr\\
    $\mathrm{log_{10}}$ Stellar Mass & [8.0,12.0] & $\mathrm{log_{10}}$ of stellar mass in solar mass ($\mathrm{M_{\odot}}$)\\
    Metallicity & [0.1,2.5] & in solar metallicity ($Z_{\odot}$) \\
    $A_V$ & [0.0,10.0] & V band dust attenuation\\
    $\eta$ & [1.,3] & Multiplicative factor producing extra attenuation for young stars in birth clouds\\
    $\mathrm{log\,U}$ & [-4.0,-2.0] & Starlight intensity on dust grains\\
    $U_{\mathrm{min}}$ & [1.0,25.0] & Lower limit of starlight intensity distribution\\
    $\gamma$ & [0.5,4.0] & Fraction of stars at $U_{\mathrm{min}}$\\
    $q_{PAH}$ & [0.01,0.99] & PAH mass fraction\\
    \hline
    \multicolumn{3}{l}{All priors are uniformly distributed in the range listed in the second column.}\\
    \multicolumn{3}{l}{The parameters in parentheses are used in independent runs to study the impact of different assumptions.}\\
    \end{tabular}
    \label{tab:SEDfitting}

\end{table*}


\end{bibunit}

\begin{thebibliography}{10}
\small
\expandafter\ifx\csname url\endcsname\relax
  \def\url#1{\texttt{#1}}\fi
\expandafter\ifx\csname urlprefix\endcsname\relax\def\urlprefix{URL }\fi
\providecommand{\bibinfo}[2]{#2}
\providecommand{\eprint}[2][]{\url{#2}}

\bibitem{Madau2014}
\bibinfo{author}{{Madau}, P.} \& \bibinfo{author}{{Dickinson}, M.}
\newblock \bibinfo{title}{{Cosmic Star-Formation History}}.
\newblock \emph{\bibinfo{journal}{\araa}} \textbf{\bibinfo{volume}{52}},
  \bibinfo{pages}{415--486} (\bibinfo{year}{2014}).

\bibitem{Bouwens15aLF}
\bibinfo{author}{{Bouwens}, R.~J.} \emph{et~al.}
\newblock \bibinfo{title}{{UV Luminosity Functions at Redshifts z \~{}4 to
  z\~{}10: 10,000 Galaxies from HST Legacy Fields}}.
\newblock \emph{\bibinfo{journal}{\apj}} \textbf{\bibinfo{volume}{803}},
  \bibinfo{pages}{34} (\bibinfo{year}{2015}).

\bibitem{Ono18}
\bibinfo{author}{{Ono}, Y.} \emph{et~al.}
\newblock \bibinfo{title}{{Great Optically Luminous Dropout Research Using
  Subaru HSC (GOLDRUSH). I. UV luminosity functions at z {\ensuremath{\sim}}
  4-7 derived with the half-million dropouts on the 100 deg$^{2}$ sky}}.
\newblock \emph{\bibinfo{journal}{\pasj}} \textbf{\bibinfo{volume}{70}},
  \bibinfo{pages}{S10} (\bibinfo{year}{2018}).

\bibitem{Watson15}
\bibinfo{author}{{Watson}, D.} \emph{et~al.}
\newblock \bibinfo{title}{{A dusty, normal galaxy in the epoch of
  reionization}}.
\newblock \emph{\bibinfo{journal}{\nat}} \textbf{\bibinfo{volume}{519}},
  \bibinfo{pages}{327--330} (\bibinfo{year}{2015}).

\bibitem{Hashimoto19}
\bibinfo{author}{{Hashimoto}, T.} \emph{et~al.}
\newblock \bibinfo{title}{{Big Three Dragons: A z = 7.15 Lyman-break galaxy
  detected in [O III] 88 {\ensuremath{\mu}}m, [C II] 158 {\ensuremath{\mu}}m,
  and dust continuum with ALMA}}.
\newblock \emph{\bibinfo{journal}{\pasj}} \textbf{\bibinfo{volume}{71}},
  \bibinfo{pages}{71} (\bibinfo{year}{2019}).

\bibitem{Tamura19}
\bibinfo{author}{{Tamura}, Y.} \emph{et~al.}
\newblock \bibinfo{title}{{Detection of the Far-infrared [O III] and Dust
  Emission in a Galaxy at Redshift 8.312: Early Metal Enrichment in the Heart
  of the Reionization Era}}.
\newblock \emph{\bibinfo{journal}{\apj}} \textbf{\bibinfo{volume}{874}},
  \bibinfo{pages}{27} (\bibinfo{year}{2019}).

\bibitem{Bakx20}
\bibinfo{author}{{Bakx}, T. J.~L.~C.} \emph{et~al.}
\newblock \bibinfo{title}{{ALMA uncovers the [C II] emission and warm dust
  continuum in a z = 8.31 Lyman break galaxy}}.
\newblock \emph{\bibinfo{journal}{\mnras}} \textbf{\bibinfo{volume}{493}},
  \bibinfo{pages}{4294--4307} (\bibinfo{year}{2020}).

\bibitem{Riechers2013}
\bibinfo{author}{{Riechers}, D.~A.} \emph{et~al.}
\newblock \bibinfo{title}{{A dust-obscured massive maximum-starburst galaxy at
  a redshift of 6.34}}.
\newblock \emph{\bibinfo{journal}{\nat}} \textbf{\bibinfo{volume}{496}},
  \bibinfo{pages}{329--333} (\bibinfo{year}{2013}).

\bibitem{Strandet17}
\bibinfo{author}{{Strandet}, M.~L.} \emph{et~al.}
\newblock \bibinfo{title}{{ISM Properties of a Massive Dusty Star-forming
  Galaxy Discovered at z {\ensuremath{\sim}} 7}}.
\newblock \emph{\bibinfo{journal}{\apjl}} \textbf{\bibinfo{volume}{842}},
  \bibinfo{pages}{L15} (\bibinfo{year}{2017}).

\bibitem{Marrone2018}
\bibinfo{author}{{Marrone}, D.~P.} \emph{et~al.}
\newblock \bibinfo{title}{{Galaxy growth in a massive halo in the first billion
  years of cosmic history}}.
\newblock \emph{\bibinfo{journal}{\nat}} \textbf{\bibinfo{volume}{553}},
  \bibinfo{pages}{51--54} (\bibinfo{year}{2018}).

\bibitem{Dudzeviciute20}
\bibinfo{author}{{Dudzevi{\v{c}}i{\={u}}t{\.{e}}}, U.} \emph{et~al.}
\newblock \bibinfo{title}{{An ALMA survey of the SCUBA-2 CLS UDS field:
  physical properties of 707 sub-millimetre galaxies}}.
\newblock \emph{\bibinfo{journal}{\mnras}} \textbf{\bibinfo{volume}{494}},
  \bibinfo{pages}{3828--3860} (\bibinfo{year}{2020}).

\bibitem{Riechers2020}
\bibinfo{author}{{Riechers}, D.~A.} \emph{et~al.}
\newblock \bibinfo{title}{{COLDz: A High Space Density of Massive Dusty
  Starburst Galaxies {\ensuremath{\sim}}1 Billion Years after the Big Bang}}.
\newblock \emph{\bibinfo{journal}{\apj}} \textbf{\bibinfo{volume}{895}},
  \bibinfo{pages}{81} (\bibinfo{year}{2020}).

\bibitem{Decarli2017}
\bibinfo{author}{{Decarli}, R.} \emph{et~al.}
\newblock \bibinfo{title}{{Rapidly star-forming galaxies adjacent to quasars at
  redshifts exceeding 6}}.
\newblock \emph{\bibinfo{journal}{\nat}} \textbf{\bibinfo{volume}{545}},
  \bibinfo{pages}{457--461} (\bibinfo{year}{2017}).

\bibitem{Mazzucchelli2019}
\bibinfo{author}{{Mazzucchelli}, C.} \emph{et~al.}
\newblock \bibinfo{title}{{Spectral Energy Distributions of Companion Galaxies
  to z {\ensuremath{\sim}} 6 Quasars}}.
\newblock \emph{\bibinfo{journal}{\apj}} \textbf{\bibinfo{volume}{881}},
  \bibinfo{pages}{163} (\bibinfo{year}{2019}).

\bibitem{Wang2019}
\bibinfo{author}{{Wang}, T.} \emph{et~al.}
\newblock \bibinfo{title}{{A dominant population of optically invisible massive
  galaxies in the early Universe}}.
\newblock \emph{\bibinfo{journal}{\nat}} \textbf{\bibinfo{volume}{572}},
  \bibinfo{pages}{211--214} (\bibinfo{year}{2019}).

\bibitem{Williams2019}
\bibinfo{author}{{Williams}, C.~C.} \emph{et~al.}
\newblock \bibinfo{title}{{Discovery of a Dark, Massive, ALMA-only Galaxy at
  z~5-6 in a Tiny 3 mm Survey}}.
\newblock \emph{\bibinfo{journal}{\apj}} \textbf{\bibinfo{volume}{884}},
  \bibinfo{pages}{154} (\bibinfo{year}{2019}).

\bibitem{Bouwens21}
\bibinfo{author}{{Bouwens}, R.~J.} \emph{et~al.}
\newblock \bibinfo{title}{{Reionization Era Bright Emission Line Survey:
  Selection and Characterization of Luminous Interstellar Medium Reservoirs in
  the z$>$6.5 Universe}}.
\newblock \emph{\bibinfo{journal}{arXiv e-prints}}
  \bibinfo{pages}{arXiv:2106.13719} (\bibinfo{year}{2021}).

\bibitem{Schreiber2015}
\bibinfo{author}{{Schreiber}, C.} \emph{et~al.}
\newblock \bibinfo{title}{{The Herschel view of the dominant mode of galaxy
  growth from z = 4 to the present day}}.
\newblock \emph{\bibinfo{journal}{\aap}} \textbf{\bibinfo{volume}{575}},
  \bibinfo{pages}{A74} (\bibinfo{year}{2015}).

\bibitem{Casey2018}
\bibinfo{author}{{Casey}, C.~M.} \emph{et~al.}
\newblock \bibinfo{title}{{The Brightest Galaxies in the Dark Ages:
  Galaxies{\textquoteright} Dust Continuum Emission during the Reionization
  Era}}.
\newblock \emph{\bibinfo{journal}{\apj}} \textbf{\bibinfo{volume}{862}},
  \bibinfo{pages}{77} (\bibinfo{year}{2018}).

\bibitem{Zavala21}
\bibinfo{author}{{Zavala}, J.~A.} \emph{et~al.}
\newblock \bibinfo{title}{{The Evolution of the IR Luminosity Function and
  Dust-obscured Star Formation over the Past 13 Billion Years}}.
\newblock \emph{\bibinfo{journal}{\apj}} \textbf{\bibinfo{volume}{909}},
  \bibinfo{pages}{165} (\bibinfo{year}{2021}).

\bibitem{Swinbank2014}
\bibinfo{author}{{Swinbank}, A.~M.} \emph{et~al.}
\newblock \bibinfo{title}{{An ALMA survey of sub-millimetre Galaxies in the
  Extended Chandra Deep Field South: the far-infrared properties of SMGs}}.
\newblock \emph{\bibinfo{journal}{\mnras}} \textbf{\bibinfo{volume}{438}},
  \bibinfo{pages}{1267--1287} (\bibinfo{year}{2014}).

\bibitem{Walter2012}
\bibinfo{author}{{Walter}, F.} \emph{et~al.}
\newblock \bibinfo{title}{{The intense starburst HDF{\,}850.1 in a galaxy
  overdensity at z{\,}{\ensuremath{\approx}}{\,}5.2 in the Hubble Deep Field}}.
\newblock \emph{\bibinfo{journal}{\nat}} \textbf{\bibinfo{volume}{486}},
  \bibinfo{pages}{233--236} (\bibinfo{year}{2012}).

\bibitem{Bowler2018}
\bibinfo{author}{{Bowler}, R.~A.~A.}, \bibinfo{author}{{Bourne}, N.},
  \bibinfo{author}{{Dunlop}, J.~S.}, \bibinfo{author}{{McLure}, R.~J.} \&
  \bibinfo{author}{{McLeod}, D.~J.}
\newblock \bibinfo{title}{{Obscured star formation in bright z~=7 Lyman-break
  galaxies}}.
\newblock \emph{\bibinfo{journal}{\mnras}} \textbf{\bibinfo{volume}{481}},
  \bibinfo{pages}{1631--1644} (\bibinfo{year}{2018}).

\end{thebibliography}


\begin{thebibliography}{10}
\small
\expandafter\ifx\csname url\endcsname\relax
  \def\url#1{\texttt{#1}}\fi
\expandafter\ifx\csname urlprefix\endcsname\relax\def\urlprefix{URL }\fi
\providecommand{\bibinfo}[2]{#2}
\providecommand{\eprint}[2][]{\url{#2}}

\bibitem{McCracken12}
\bibinfo{author}{{McCracken}, H.~J.} \emph{et~al.}
\newblock \bibinfo{title}{{UltraVISTA: a new ultra-deep near-infrared survey in
  COSMOS}}.
\newblock \emph{\bibinfo{journal}{\aap}} \textbf{\bibinfo{volume}{544}},
  \bibinfo{pages}{A156} (\bibinfo{year}{2012}).

\bibitem{Jarvis13}
\bibinfo{author}{{Jarvis}, M.~J.} \emph{et~al.}
\newblock \bibinfo{title}{{The VISTA Deep Extragalactic Observations (VIDEO)
  survey}}.
\newblock \emph{\bibinfo{journal}{\mnras}} \textbf{\bibinfo{volume}{428}},
  \bibinfo{pages}{1281--1295} (\bibinfo{year}{2013}).

\bibitem{Erben09}
\bibinfo{author}{{Erben}, T.} \emph{et~al.}
\newblock \bibinfo{title}{{CARS: the CFHTLS-Archive-Research Survey. I.
  Five-band multi-colour data from 37 sq. deg. CFHTLS-wide observations}}.
\newblock \emph{\bibinfo{journal}{\aap}} \textbf{\bibinfo{volume}{493}},
  \bibinfo{pages}{1197--1222} (\bibinfo{year}{2009}).

\bibitem{Aihara18}
\bibinfo{author}{{Aihara}, H.} \emph{et~al.}
\newblock \bibinfo{title}{{First data release of the Hyper Suprime-Cam Subaru
  Strategic Program}}.
\newblock \emph{\bibinfo{journal}{\pasj}} \textbf{\bibinfo{volume}{70}},
  \bibinfo{pages}{S8} (\bibinfo{year}{2018}).

\bibitem{Bowler20}
\bibinfo{author}{{Bowler}, R.~A.~A.} \emph{et~al.}
\newblock \bibinfo{title}{{A lack of evolution in the very bright end of the
  galaxy luminosity function from z ~= 8 to 10}}.
\newblock \emph{\bibinfo{journal}{\mnras}} \textbf{\bibinfo{volume}{493}},
  \bibinfo{pages}{2059--2084} (\bibinfo{year}{2020}).

\bibitem{Stefanon19}
\bibinfo{author}{{Stefanon}, M.} \emph{et~al.}
\newblock \bibinfo{title}{{The Brightest z {\ensuremath{\gtrsim}} 8 Galaxies
  over the COSMOS UltraVISTA Field}}.
\newblock \emph{\bibinfo{journal}{\apj}} \textbf{\bibinfo{volume}{883}},
  \bibinfo{pages}{99} (\bibinfo{year}{2019}).

\bibitem{Bowler17}
\bibinfo{author}{{Bowler}, R.~A.~A.}, \bibinfo{author}{{Dunlop}, J.~S.},
  \bibinfo{author}{{McLure}, R.~J.} \& \bibinfo{author}{{McLeod}, D.~J.}
\newblock \bibinfo{title}{{Unveiling the nature of bright z ~= 7 galaxies with
  the Hubble Space Telescope}}.
\newblock \emph{\bibinfo{journal}{\mnras}} \textbf{\bibinfo{volume}{466}},
  \bibinfo{pages}{3612--3635} (\bibinfo{year}{2017}).

\bibitem{Schaerer2020}
\bibinfo{author}{{Schaerer}, D.} \emph{et~al.}
\newblock \bibinfo{title}{{The ALPINE-ALMA [C II] survey. Little to no
  evolution in the [C II]-SFR relation over the last 13 Gyr}}.
\newblock \emph{\bibinfo{journal}{\aap}} \textbf{\bibinfo{volume}{643}},
  \bibinfo{pages}{A3} (\bibinfo{year}{2020}).

\bibitem{DeLooze2014}
\bibinfo{author}{{De Looze}, I.} \emph{et~al.}
\newblock \bibinfo{title}{{The applicability of far-infrared fine-structure
  lines as star formation rate tracers over wide ranges of metallicities and
  galaxy types}}.
\newblock \emph{\bibinfo{journal}{\aap}} \textbf{\bibinfo{volume}{568}},
  \bibinfo{pages}{A62} (\bibinfo{year}{2014}).

\bibitem{Carnall2018}
\bibinfo{author}{{Carnall}, A.~C.}, \bibinfo{author}{{McLure}, R.~J.},
  \bibinfo{author}{{Dunlop}, J.~S.} \& \bibinfo{author}{{Dav{\'e}}, R.}
\newblock \bibinfo{title}{{Inferring the star formation histories of massive
  quiescent galaxies with BAGPIPES: evidence for multiple quenching
  mechanisms}}.
\newblock \emph{\bibinfo{journal}{\mnras}} \textbf{\bibinfo{volume}{480}},
  \bibinfo{pages}{4379--4401} (\bibinfo{year}{2018}).

\bibitem{Bruzual03}
\bibinfo{author}{{Bruzual}, G.} \& \bibinfo{author}{{Charlot}, S.}
\newblock \bibinfo{title}{{Stellar population synthesis at the resolution of
  2003}}.
\newblock \emph{\bibinfo{journal}{\mnras}} \textbf{\bibinfo{volume}{344}},
  \bibinfo{pages}{1000--1028} (\bibinfo{year}{2003}).

\bibitem{Kroupa02}
\bibinfo{author}{{Kroupa}, P.} \& \bibinfo{author}{{Boily}, C.~M.}
\newblock \bibinfo{title}{{On the mass function of star clusters}}.
\newblock \emph{\bibinfo{journal}{\mnras}} \textbf{\bibinfo{volume}{336}},
  \bibinfo{pages}{1188--1194} (\bibinfo{year}{2002}).

\bibitem{Byler17}
\bibinfo{author}{{Byler}, N.}, \bibinfo{author}{{Dalcanton}, J.~J.},
  \bibinfo{author}{{Conroy}, C.} \& \bibinfo{author}{{Johnson}, B.~D.}
\newblock \bibinfo{title}{{Nebular Continuum and Line Emission in Stellar
  Population Synthesis Models}}.
\newblock \emph{\bibinfo{journal}{\apj}} \textbf{\bibinfo{volume}{840}},
  \bibinfo{pages}{44} (\bibinfo{year}{2017}).

\bibitem{Ferland17}
\bibinfo{author}{{Ferland}, G.~J.} \emph{et~al.}
\newblock \bibinfo{title}{{The 2017 Release Cloudy}}.
\newblock \emph{\bibinfo{journal}{\rmxaa}} \textbf{\bibinfo{volume}{53}},
  \bibinfo{pages}{385--438} (\bibinfo{year}{2017}).

\bibitem{Calzetti00}
\bibinfo{author}{{Calzetti}, D.} \emph{et~al.}
\newblock \bibinfo{title}{{The Dust Content and Opacity of Actively
  Star-forming Galaxies}}.
\newblock \emph{\bibinfo{journal}{\apj}} \textbf{\bibinfo{volume}{533}},
  \bibinfo{pages}{682--695} (\bibinfo{year}{2000}).

\bibitem{Charlot2000}
\bibinfo{author}{{Charlot}, S.} \& \bibinfo{author}{{Fall}, S.~M.}
\newblock \bibinfo{title}{{A Simple Model for the Absorption of Starlight by
  Dust in Galaxies}}.
\newblock \emph{\bibinfo{journal}{\apj}} \textbf{\bibinfo{volume}{539}},
  \bibinfo{pages}{718--731} (\bibinfo{year}{2000}).

\bibitem{DrainLi07}
\bibinfo{author}{{Draine}, B.~T.} \& \bibinfo{author}{{Li}, A.}
\newblock \bibinfo{title}{{Infrared Emission from Interstellar Dust. IV. The
  Silicate-Graphite-PAH Model in the Post-Spitzer Era}}.
\newblock \emph{\bibinfo{journal}{\apj}} \textbf{\bibinfo{volume}{657}},
  \bibinfo{pages}{810--837} (\bibinfo{year}{2007}).

\bibitem{Wang2013}
\bibinfo{author}{{Wang}, R.} \emph{et~al.}
\newblock \bibinfo{title}{{Star Formation and Gas Kinematics of Quasar Host
  Galaxies at z \raisebox{-0.5ex}\textasciitilde 6: New Insights from ALMA}}.
\newblock \emph{\bibinfo{journal}{\apj}} \textbf{\bibinfo{volume}{773}},
  \bibinfo{pages}{44} (\bibinfo{year}{2013}).

\bibitem{Capak15}
\bibinfo{author}{{Capak}, P.~L.} \emph{et~al.}
\newblock \bibinfo{title}{{Galaxies at redshifts 5 to 6 with systematically low
  dust content and high [C II] emission}}.
\newblock \emph{\bibinfo{journal}{\nat}} \textbf{\bibinfo{volume}{522}},
  \bibinfo{pages}{455--458} (\bibinfo{year}{2015}).

\bibitem{Dessauges2020}
\bibinfo{author}{{Dessauges-Zavadsky}, M.} \emph{et~al.}
\newblock \bibinfo{title}{{The ALPINE-ALMA [C II] survey. Molecular gas budget
  in the early Universe as traced by [C II]}}.
\newblock \emph{\bibinfo{journal}{\aap}} \textbf{\bibinfo{volume}{643}},
  \bibinfo{pages}{A5} (\bibinfo{year}{2020}).

\bibitem{Casey2012}
\bibinfo{author}{{Casey}, C.~M.}
\newblock \bibinfo{title}{{Far-infrared spectral energy distribution fitting
  for galaxies near and far}}.
\newblock \emph{\bibinfo{journal}{\mnras}} \textbf{\bibinfo{volume}{425}},
  \bibinfo{pages}{3094--3103} (\bibinfo{year}{2012}).

\bibitem{Schreiber18}
\bibinfo{author}{{Schreiber}, C.} \emph{et~al.}
\newblock \bibinfo{title}{{Dust temperature and mid-to-total infrared color
  distributions for star-forming galaxies at $0 < z < 4$}}.
\newblock \emph{\bibinfo{journal}{\aap}} \textbf{\bibinfo{volume}{609}},
  \bibinfo{pages}{A30} (\bibinfo{year}{2018}).

\bibitem{Faisst2020}
\bibinfo{author}{{Faisst}, A.~L.} \emph{et~al.}
\newblock \bibinfo{title}{{ALMA Characterises the Dust Temperature of z
  \raisebox{-0.5ex}\textasciitilde 5.5 Star-Forming Galaxies}}.
\newblock \emph{\bibinfo{journal}{arXiv e-prints}}
  \bibinfo{pages}{arXiv:2005.07716} (\bibinfo{year}{2020}).

\bibitem{Dacunha2013}
\bibinfo{author}{{da Cunha}, E.} \emph{et~al.}
\newblock \bibinfo{title}{{On the Effect of the Cosmic Microwave Background in
  High-redshift (Sub-)millimeter Observations}}.
\newblock \emph{\bibinfo{journal}{\apj}} \textbf{\bibinfo{volume}{766}},
  \bibinfo{pages}{13} (\bibinfo{year}{2013}).

\bibitem{Laporte17}
\bibinfo{author}{{Laporte}, N.} \emph{et~al.}
\newblock \bibinfo{title}{{Dust in the Reionization Era: ALMA Observations of a
  z = 8.38 Gravitationally Lensed Galaxy}}.
\newblock \emph{\bibinfo{journal}{\apjl}} \textbf{\bibinfo{volume}{837}},
  \bibinfo{pages}{L21} (\bibinfo{year}{2017}).

\bibitem{Behrens2018}
\bibinfo{author}{{Behrens}, C.}, \bibinfo{author}{{Pallottini}, A.},
  \bibinfo{author}{{Ferrara}, A.}, \bibinfo{author}{{Gallerani}, S.} \&
  \bibinfo{author}{{Vallini}, L.}
\newblock \bibinfo{title}{{Dusty galaxies in the Epoch of Reionization:
  simulations}}.
\newblock \emph{\bibinfo{journal}{\mnras}} \textbf{\bibinfo{volume}{477}},
  \bibinfo{pages}{552--565} (\bibinfo{year}{2018}).

\bibitem{Liang2019}
\bibinfo{author}{{Liang}, L.} \emph{et~al.}
\newblock \bibinfo{title}{{On the dust temperatures of high-redshift
  galaxies}}.
\newblock \emph{\bibinfo{journal}{\mnras}} \textbf{\bibinfo{volume}{489}},
  \bibinfo{pages}{1397--1422} (\bibinfo{year}{2019}).

\bibitem{Sommovigo2020}
\bibinfo{author}{{Sommovigo}, L.} \emph{et~al.}
\newblock \bibinfo{title}{{Warm dust in high-z galaxies: origin and
  implications}}.
\newblock \emph{\bibinfo{journal}{arXiv e-prints}}
  \bibinfo{pages}{arXiv:2004.09528} (\bibinfo{year}{2020}).

\bibitem{deVis19}
\bibinfo{author}{{De Vis}, P.} \emph{et~al.}
\newblock \bibinfo{title}{{A systematic metallicity study of DustPedia galaxies
  reveals evolution in the dust-to-metal ratios}}.
\newblock \emph{\bibinfo{journal}{\aap}} \textbf{\bibinfo{volume}{623}},
  \bibinfo{pages}{A5} (\bibinfo{year}{2019}).

\bibitem{Mancini16}
\bibinfo{author}{{Mancini}, M.} \emph{et~al.}
\newblock \bibinfo{title}{{Interpreting the evolution of galaxy colours from z
  = 8 to 5}}.
\newblock \emph{\bibinfo{journal}{\mnras}} \textbf{\bibinfo{volume}{462}},
  \bibinfo{pages}{3130--3145} (\bibinfo{year}{2016}).

\bibitem{Graziani20}
\bibinfo{author}{{Graziani}, L.} \emph{et~al.}
\newblock \bibinfo{title}{{The assembly of dusty galaxies at z
  {\ensuremath{\geq}} 4: statistical properties}}.
\newblock \emph{\bibinfo{journal}{\mnras}} \textbf{\bibinfo{volume}{494}},
  \bibinfo{pages}{1071--1088} (\bibinfo{year}{2020}).

\bibitem{Gruppioni13}
\bibinfo{author}{{Gruppioni}, C.} \emph{et~al.}
\newblock \bibinfo{title}{{The Herschel PEP/HerMES luminosity function - I.
  Probing the evolution of PACS selected Galaxies to z ~= 4}}.
\newblock \emph{\bibinfo{journal}{\mnras}} \textbf{\bibinfo{volume}{432}},
  \bibinfo{pages}{23--52} (\bibinfo{year}{2013}).

\bibitem{Carilli13}
\bibinfo{author}{{Carilli}, C.~L.} \& \bibinfo{author}{{Walter}, F.}
\newblock \bibinfo{title}{{Cool Gas in High-Redshift Galaxies}}.
\newblock \emph{\bibinfo{journal}{\araa}} \textbf{\bibinfo{volume}{51}},
  \bibinfo{pages}{105--161} (\bibinfo{year}{2013}).

\bibitem{Peebles80}
\bibinfo{author}{{Peebles}, P.~J.~E.}
\newblock \emph{\bibinfo{title}{{The large-scale structure of the universe}}}
  (\bibinfo{year}{1980}).

\bibitem{BaroneNugent14}
\bibinfo{author}{{Barone-Nugent}, R.~L.} \emph{et~al.}
\newblock \bibinfo{title}{{Measurement of Galaxy Clustering at z \~{} 7.2 and
  the Evolution of Galaxy Bias from 3.8 {\lt} z {\lt} 8 in the XDF, GOODS-S,
  and GOODS-N}}.
\newblock \emph{\bibinfo{journal}{\apj}} \textbf{\bibinfo{volume}{793}},
  \bibinfo{pages}{17} (\bibinfo{year}{2014}).

\bibitem{Adelberger05}
\bibinfo{author}{{Adelberger}, K.~L.} \emph{et~al.}
\newblock \bibinfo{title}{{The Spatial Clustering of Star-forming Galaxies at
  Redshifts 1.4 $<$ z $<$ 3.5}}.
\newblock \emph{\bibinfo{journal}{\apj}} \textbf{\bibinfo{volume}{619}},
  \bibinfo{pages}{697--713} (\bibinfo{year}{2005}).

\bibitem{Qiu18}
\bibinfo{author}{{Qiu}, Y.} \emph{et~al.}
\newblock \bibinfo{title}{{Dependence of galaxy clustering on UV luminosity and
  stellar mass at z {\ensuremath{\sim}} 4-7}}.
\newblock \emph{\bibinfo{journal}{\mnras}} \textbf{\bibinfo{volume}{481}},
  \bibinfo{pages}{4885--4894} (\bibinfo{year}{2018}).

\bibitem{Bhowmick20}
\bibinfo{author}{{Bhowmick}, A.~K.} \emph{et~al.}
\newblock \bibinfo{title}{{Cosmic variance of z $>$ 7 galaxies: prediction from
  BLUETIDES}}.
\newblock \emph{\bibinfo{journal}{\mnras}} \textbf{\bibinfo{volume}{496}},
  \bibinfo{pages}{754--766} (\bibinfo{year}{2020}).

\bibitem{Uzgil21}
\bibinfo{author}{{Uzgil}, B.~D.} \emph{et~al.}
\newblock \bibinfo{title}{{The ALMA Spectroscopic Survey in the HUDF: A Search
  for [C II] Emitters at 6 {\ensuremath{\leq}} z {\ensuremath{\leq}} 8}}.
\newblock \emph{\bibinfo{journal}{\apj}} \textbf{\bibinfo{volume}{912}},
  \bibinfo{pages}{67} (\bibinfo{year}{2021}).

\bibitem{Whitaker2017}
\bibinfo{author}{{Whitaker}, K.~E.} \emph{et~al.}
\newblock \bibinfo{title}{{The Constant Average Relationship between
  Dust-obscured Star Formation and Stellar Mass from z = 0 to z = 2.5}}.
\newblock \emph{\bibinfo{journal}{\apj}} \textbf{\bibinfo{volume}{850}},
  \bibinfo{pages}{208} (\bibinfo{year}{2017}).

\bibitem{Fudamoto2020}
\bibinfo{author}{{Fudamoto}, Y.} \emph{et~al.}
\newblock \bibinfo{title}{{The ALPINE-ALMA [CII] survey. Dust attenuation
  properties and obscured star formation at z {\ensuremath{\sim}} 4.4-5.8}}.
\newblock \emph{\bibinfo{journal}{\aap}} \textbf{\bibinfo{volume}{643}},
  \bibinfo{pages}{A4} (\bibinfo{year}{2020}).

\bibitem{Bethermin15}
\bibinfo{author}{{B{\'e}thermin}, M.} \emph{et~al.}
\newblock \bibinfo{title}{{Evolution of the dust emission of massive galaxies
  up to z = 4 and constraints on their dominant mode of star formation}}.
\newblock \emph{\bibinfo{journal}{\aap}} \textbf{\bibinfo{volume}{573}},
  \bibinfo{pages}{A113} (\bibinfo{year}{2015}).

\bibitem{Scoville07a}
\bibinfo{author}{{Scoville}, N.} \emph{et~al.}
\newblock \bibinfo{title}{{COSMOS: Hubble Space Telescope Observations}}.
\newblock \emph{\bibinfo{journal}{\apjs}} \textbf{\bibinfo{volume}{172}},
  \bibinfo{pages}{38--45} (\bibinfo{year}{2007}).

\end{thebibliography}
\end{document}